\documentclass[%
 reprint,
 amsmath,amssymb,
 aps,
pre,
]{revtex4-1}

\usepackage{graphicx,amssymb,amsfonts,amsmath}

\usepackage{graphicx,bm}
\usepackage{epsfig}
\usepackage{color}
\usepackage{epsfig,dsfont,amssymb,amsmath,amsthm,amsfonts,amsbsy,mathrsfs}

\usepackage{ulem}

\usepackage[utf8]{inputenc}

\usepackage{color}
\definecolor{darkblue}{rgb}{0,0,0.6}
\definecolor{darkred}{rgb}{0.6,0,0}
\definecolor{darkgreen}{rgb}{0,0.6,0}
\definecolor{purple}{rgb}{0.6,0,0.6}
\definecolor{gray}{rgb}{0.5,0.5,0.5}

\newif\iffigures
\figurestrue 

\newif\ifcomments
\commentstrue 

\ifcomments
 \newcommand{\ABKq}[1]{\textcolor{blue}{ABK:#1}}
 \newcommand{\EEFq}[1]{\textcolor{purple}{EEF:#1}}
 \newcommand{\ARq}[1]{\textcolor{magenta}{AR:#1}}
 \newcommand{\LFq}[1]{\textcolor{darkgreen}{LF:#1}}
 \newcommand{\TGq}[1]{\textcolor{red}{TG:#1}}
\else
 \newcommand{\ABKq}[1]{}
 \newcommand{\EEFq}[1]{}
 \newcommand{\ARq}[1]{}
 \newcommand{\LFq}[1]{}
 \newcommand{\TGq}[1]{}
\fi

\newcommand{\Lopt}{{L}_{\text{opt}}} 

\newcommand{\Leve}{{L}_{\text{eve}}}
\newcommand{\Seve}{{S}_{\text{eve}}}
 
\newcommand{\Lc}{{{L}_{\text{c}}}}

\newcommand{\Sclus}{{S}_{\text{clust}}}
\newcommand{\Sopt}{{S}_{\text{opt}}}

\newcommand{\hopt}{{h}_{\text{opt}}} 

\newcommand{\Sact}{{S}_{\text{act}}}

\newcommand{\dep}{_\text{dep}}

\newcommand{\zetadep}{{\zeta_{\text{dep}}}}
\newcommand{\zetaeq}{{\zeta_{\text{eq}}}}

\newcommand{\zetaeqRF}{{\zeta^{\text{RF}}_{\text{eq}}}}
\newcommand{\taudep}{{\tau_{\text{dep}}}}
\newcommand{\taueq}{{\tau_{\text{eq}}}}
\newcommand{\taueqRF}{{\tau^{\text{RF}}_{\text{eq}}}}
\newcommand{\nudep}{{\nu_{\text{dep}}}}

\newcommand{\nueq}{{\nu_{\text{eq}}}}
\newcommand{\nueqRF}{{\nu^{\text{RF}}_{\text{eq}}}}
\newcommand{\muRF}{{\mu^{\text{RF}}}}
\newcommand{\thetaeq}{{\theta_{\text{eq}}}}
\newcommand{\thetaeqRF}{{\theta^{\text RF}_{\text{eq}}}}
\newcommand{\betadep}{{\beta_{\text{dep}}}}

\newcommand{\fc}{{f_c}}
\newcommand{\pinningforce}{{F_{\text{p}}}}

\begin{document}

\title{Spatio-temporal patterns in ultra-slow domain wall creep dynamics}

\author{Ezequiel E. Ferrero}
\affiliation{Universit\'e Grenoble Alpes, LIPHY, F-38000 Grenoble, France and CNRS, LIPHY, F-38000 Grenoble, France}
\author{Laura Foini}
\affiliation{Department of Quantum Matter Physics, University of Geneva, 24 Quai Ernest-Ansermet, CH-1211 Geneva, Switzerland}
\author{Thierry Giamarchi}
\affiliation{Department of Quantum Matter Physics, University of Geneva, 24 Quai Ernest-Ansermet, CH-1211 Geneva, Switzerland}
\author{Alejandro B. Kolton}
\affiliation{Instituto Balseiro-UNCu and CONICET, Centro Atómico Bariloche, 8400 Bariloche, Argentina}
\author{Alberto Rosso}
\affiliation{LPTMS, CNRS, Univ. Paris-Sud, Université Paris-Saclay, 91405 Orsay, France}

\date{\today}

\begin{abstract}
 In presence of impurities, ferromagnetic and ferroelectric domain walls slide only above a finite external field.
 Close to this depinning threshold, they proceed by large and abrupt jumps, called avalanches, while, at much smaller field,
 these interfaces creep by thermal activation.
 In this work we develop a novel numerical technique that captures the ultra-slow creep regime over huge time scales. 
 We point out the existence of activated events that involve collective reorganizations similar to avalanches, but,
 at variance with them, display correlated spatio-temporal patterns that resemble the complex sequence of aftershocks
 observed after a large earthquake.
 Remarkably, we show that events assembly in independent clusters that display at large scales the same statistics as
 critical depinning avalanches. 
 We foresee this correlated dynamics being experimentally accessible by magneto-optical imaging of ferromagnetic films. 
\end{abstract}


\maketitle

The physics of disordered elastic systems (DES) is relevant for many areas of physics 
such as magnetic~\cite{Lemerle-PRL1998,YCMDO06,MJMCF07,RBJFM04,Jeudy2016} and
ferroelectric~\cite{TPGT02,PT06} domain-wall, contact lines in wetting~\cite{MGR2002},
crack propagation~\cite{BBFRR2002,ZAN2006} and vortex lines in type-II superconductors~\cite{BFGLV94}.
It involves the driven motion of an elastic object, such as a manifold or a periodic structure,
in a weakly disordered medium.
At zero temperature, setting the system in motion requires to apply a finite force $f$ 
exceeding a critical value $f_c$, a process known as depinning.
When $f \gg f_c$, and due to dissipation, the system flows with a velocity essentially
proportional to the driving force $f$, while it is pinned for $f<\fc$.
At finite temperature this behavior is drastically modified since energy barriers can always
be passed by thermal activation, leading to a finite velocity for any finite force. 

One of the important questions is the response of an elastic interface to a very
small force, $f \ll \fc$.
Understanding this regime is relevant to assess the electrical resistance of
a superconductor~\cite{BFGLV94}, or to judge in which conditions ferroelectric or 
ferromagnetic materials can be used to store and retrieve
information~\cite{Parkin-Science2008}.
It is now well known that at very low driving the response is highly non-linear,
leading to the so-called {\it creep regime}.
Phenomenological arguments based on the Arrhenius activation of segments of
the interface (thermal nuclei) showed that, instead of a linear response,
one should expect a stretched exponential
response~\cite{Ioffe_Vinokur_1987,NattermannEPL1987,Feigelman1989,NattermannPRB1990,Dong1993},
where the average velocity of the wall is exponentially small in a power-law of the external force.
This behavior was later confirmed by more microscopic derivations 
based on a functional renormalization group procedure (FRG) in $d=4-\epsilon$ 
dimensions~\cite{CGLD98,Chauve-PRB2000}.
Experiments on magnetic and ferroelectric domain walls also provided
confirmation of the creep law for the average velocity~\cite{Lemerle-PRL1998,TPGT02}.

Despite these important success between experiments and FRG,
two important questions remain open concerning the creep motion:
First, a convincing confirmation of the creep law.
In fact, while experiments typically concern systems in $d=1$ and $d=2$,
the FRG is valid only in $d=4-\epsilon$, with $\epsilon$ assumed to be small.
The only available theoretical tools to address the dynamics of low dimensional interfaces,
so far, are numerical simulations.
In this respect, traditional molecular dynamics techniques have difficulties reaching the
very long times which are necessary to deal with the ultra-slow motion characterizing creep.
Thus, a well controlled numerical technique that would not suffer from slowing down when
the force is reduced would be highly suitable.
Second, and most important, the understanding of creep dynamics besides its mean velocity
is an open issue.
The FRG suggests that beyond the size of the thermal nucleus the coarse-grained 
motion should be depinning-like
up to a second temperature-controlled and very large length scale where the flow regime
occurs~\cite{CGLD98,Chauve-PRB2000}.
However, the evidence of such large scale and the description of this depinning-like motion
are still elusive.

In this paper we provide a novel numerical technique able to tackle the creep at very small
forces and vanishing temperature by computing the full sequence of activated events.
We determine the distribution of their sizes -- which displays an anomalous power law behavior --
and study their surprising  spatio-temporal organization: 
Initial seeds trigger large reorganizations of events (as shown in Fig. \ref{fig:Patterns}),
statistically identical at large scales to deterministic depinning avalanches.
In this way, our results give a clear interpretation of the FRG predictions.
Further, they link the creep motion with the complex earthquake dynamics~\cite{Scholz_2002}
where a main shock triggers a cascade of aftershocks \cite{JLR_2014,JK_2010}.
We foresee this correlated dynamics being experimentally accessible by events 
detection using magneto-optical imaging of ferromagnetic films \cite{RBJFM04}. 
To motivate such experimental test we provide the relevant 
scales for the particular case of Pt/Co/Pt thin 
ferromagnetic films~\cite{GBFJKG14}.

\iffigures
\begin{figure}
\includegraphics[width=\columnwidth]{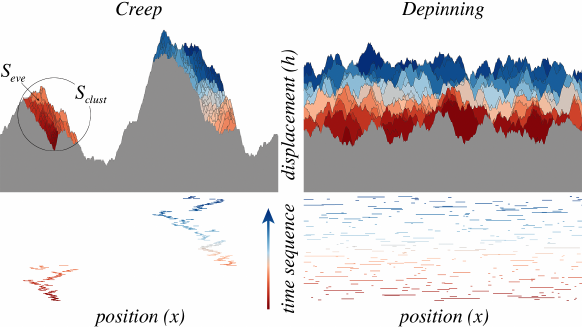}
\caption{\label{fig:Patterns}
  \textit{Spatio-temporal patterns~--} 
  Top:
  Snapshot of $300$ consecutive configurations for the moving interface.
  On the left, a typical sequence of events at a small force 
  in the creep regime that assemble in space on a pattern of two clusters.
  On the right, a typical sequence of deterministic avalanches at a larger force 
  close to depinning, that appear randomly distributed in space.
  Bottom:
  Activity maps showing for each event of the top row an horizontal segment with 
  its lateral extension.
  In all cases the time sequence is illustrated by a color code, from dark red (older) 
  to dark blue (more recent). 
  }
\end{figure}

\smallskip \noindent {\it Phenomenology.}
We consider a $d=1$-dimensional interface in absence of overhangs.
At any time $t$, the local displacement is described by a single valued function, $h(x,t)$,
which in the overdamped limit is described by the so-called 
quenched Edwards-Wilkinson equation~\cite{Fisher_1998,Kardar-PR1998,FBKR13}:
\begin{equation}
\label{equation1}
\gamma \partial_t h(x,t)= c\nabla^2 h(x,t) + f + \pinningforce(x,h)+\eta(x,t)
\end{equation}
where $c\nabla^2 h(x,t)$ accounts for the elastic force due to the surface tension, 
$f$ is the external pulling force, and the fluctuations
induced by impurities and temperature are encoded in the quenched disorder term 
$\pinningforce(x,h)$ and in the Langevin thermal noise $\eta(x,t)$ respectively.
We consider here Random Bond (RB) disorder, in which
the pinning potential is short-range correlated in the direction of motion.
The analysis of the Random-Field disorder case, where the pinning force is
short-range correlated, is deferred to the Supplemental Material (SM).

\iffigures
 \begin{figure}[ht]
 \includegraphics[width=0.65\columnwidth]{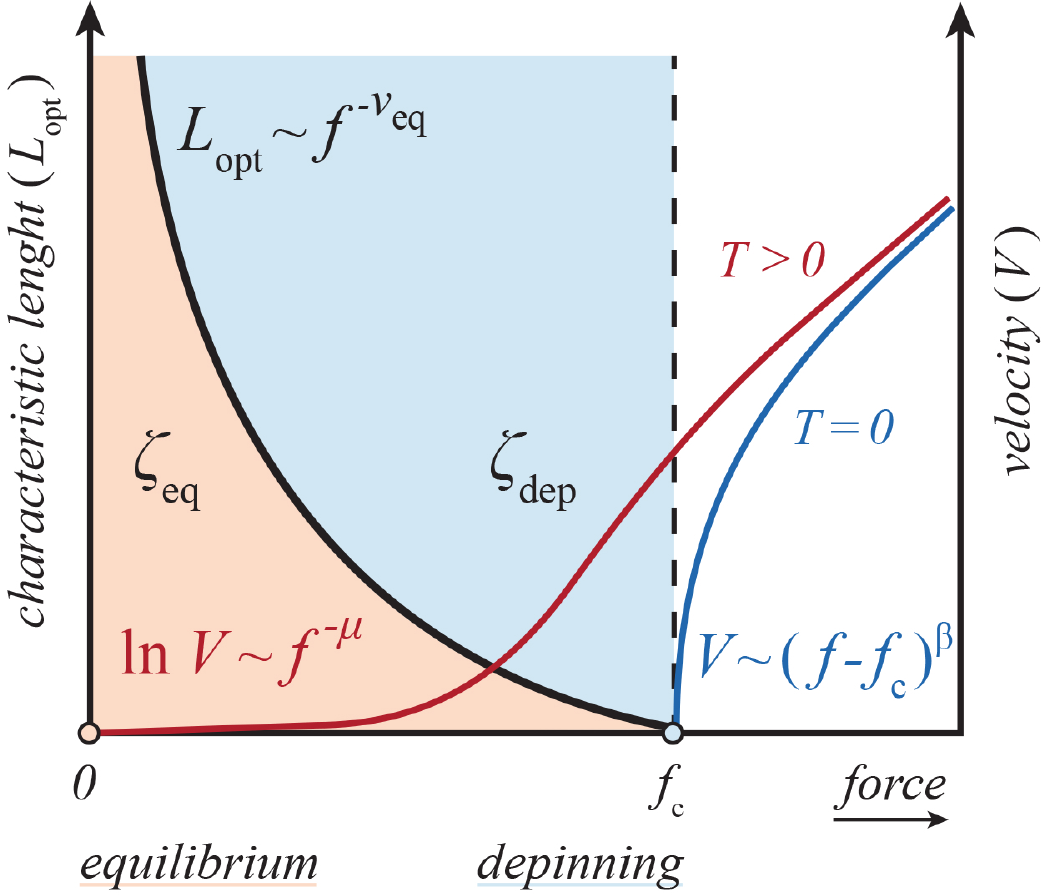} 
   \caption{\label{fig:VFchar_and_Schem}
   \textit{Velocity-force characteristics and reference points~--}
    Two aspects of the $T>0$ dynamics below $\fc$ are schematically shown:
    (i) The velocity $V$ has a finite value at low forces well approximated by Eq.~\ref{creepformula}, and
    (ii) A characteristic length scale $\Lopt$, that diverges as $f$ goes to zero, separates
    two dynamical regimes identified by different roughness exponents, $\zetaeq$ below $\Lopt$ and
    $\zetadep$ above $\Lopt$.
   }
 \end{figure}
\fi

At zero temperature there are two reference points where self-affinity and scale-free behavior are expected: 

{\em Equilibrium scaling.}
The first, at $f=0$, corresponds to thermodynamic equilibrium where the extensive ground state energy displays critical
sample to sample fluctuations, growing as $L^{\thetaeq}$ and the domain wall is rough with a self-affine width growing
as $L^{\zetaeq}$.
The exponents $\thetaeq$ and $\zetaeq$ are universal and depend only on dimension,
range of elastic interactions and disorder class.

{\em Depinning scaling.}
The second critical point, at $f=\fc$ and zero temperature, corresponds to the depinning transition above which
the interface acquires a finite global velocity $V\sim (f-\fc)^\betadep$.
This point is characterized by a roughness $\zetadep$ (see Fig. \ref{fig:VFchar_and_Schem} at $f=\fc$).
At any force close to $\fc$ a small perturbation can induce a large reorganization of the interface,
called depinning avalanche.
The avalanche size, namely the area spanned by the moving interface, has power-law statistics
with exponent~\cite{NF93,RLDW09}:
\begin{equation}\label{Eq_tau}
 \tau = \taudep = 2 - \frac{2}{d+\zetadep} 
\end{equation}

At small but {\it finite} temperature ($T>0$) and below $\fc$ the instantaneous dynamics appears
as a collection of incoherent vibrations localized around deep metastable configurations.
However, the presence of a small positive drive makes a global forward motion energetically favorable
in the long term and the steady-state dynamics is well described by the {\em creep scaling},
which predicts a ``typical" size for the thermal activated nucleus:
\begin{equation}
\label{eq:Lopt}
\Lopt(f) \sim 1/f^{\nueq} \quad \text{with} \quad \nueq = \frac{1}{d+\zetaeq-\thetaeq} \ .
\end{equation}

The global velocity is determined using the Arrhenius formula, assuming
that the energy barriers scale as the ground state fluctuations, i.e. as $\Lopt^{\theta}$ with $\theta=\thetaeq$,
resulting in the creep law~\cite{Ioffe_Vinokur_1987,NattermannEPL1987,Feigelman1989,NattermannPRB1990,Dong1993}:
\begin{equation} \label{creepformula}
- \log V \propto  f^{-\mu} \ ,
\end{equation}
where $\mu=\nueq\theta=\theta/(d+\zetaeq-\thetaeq)$.

\smallskip \noindent {\it Modeling.}
Traditional integration schemes for Eq.\ref{equation1}, like molecular dynamics,
fail to capture the long time scales associated to the activated events.
Fortunately, it was shown that, at vanishing temperature ($T=0^+$), it is
possible to target the rare events that move the interface forward irreversibly
and build a dynamics based on them~\cite{Kolton-PRL2006,Kolton-PRB2009}.
This method describes the limit in which motion is effectively dominated 
by a sequence of metastable states of decreasing energy separated by energy
barriers that govern the waiting time among jumps.
The choice of the next metastable state amounts to find the \textit{minimal}
energy barrier to be overcome in order to reach a state with smaller energy.
This corresponds to enumerate all pathways that end in a state with lower
energy and select the one that has overcome the smallest barrier.
Unfortunately, this protocol is very expensive: its computing time grows
exponentially with $\Lopt$.
 
In order to explore the low force regime, where creep scaling laws are expected to apply
and $\Lopt$ is very large, we adopt here a \textit{new strategy}.
In few words, our approach consists in selecting the minimal rearrangement \textit{in size}
that takes the interface to a lower energy state, avoiding the exponential cost in $\Lopt$.
At small forces, this approximated method is equivalent to search for a minimal barrier
since they grow extensively with the size of the rearrangement. 
We refer the reader to the SM for a validation of our approximation and algorithmic details.

The dynamics that evolves the interface from one metastable state to the next one 
is composed by two steps:
an \textit{activated move} to jump a barrier and find a lower energy state,
followed by a \textit{deterministic relaxation}, that further drives the interface
through the energy lowering gradient till the next minimum is reached.
The difference between the new and the previous metastable configurations
is a compact object that we call {\it event}, well characterized both by its area
$\Seve$ and by its lateral size $\Leve$ (equal to the number of monomers involved in the jump). 
For a given realization of the disorder, the sequence of metastable configurations generated with 
our algorithm is unique once the steady-state is reached.
A typical sequence of locked configurations can be seen in Fig. \ref{fig:Patterns} (left). 
Unless specified, all the reported numerical data correspond to a system size $L=3360$.

\smallskip \noindent {\it Results analysis.}
\iffigures
 \begin{figure}
\center
\includegraphics[width=0.93\columnwidth]{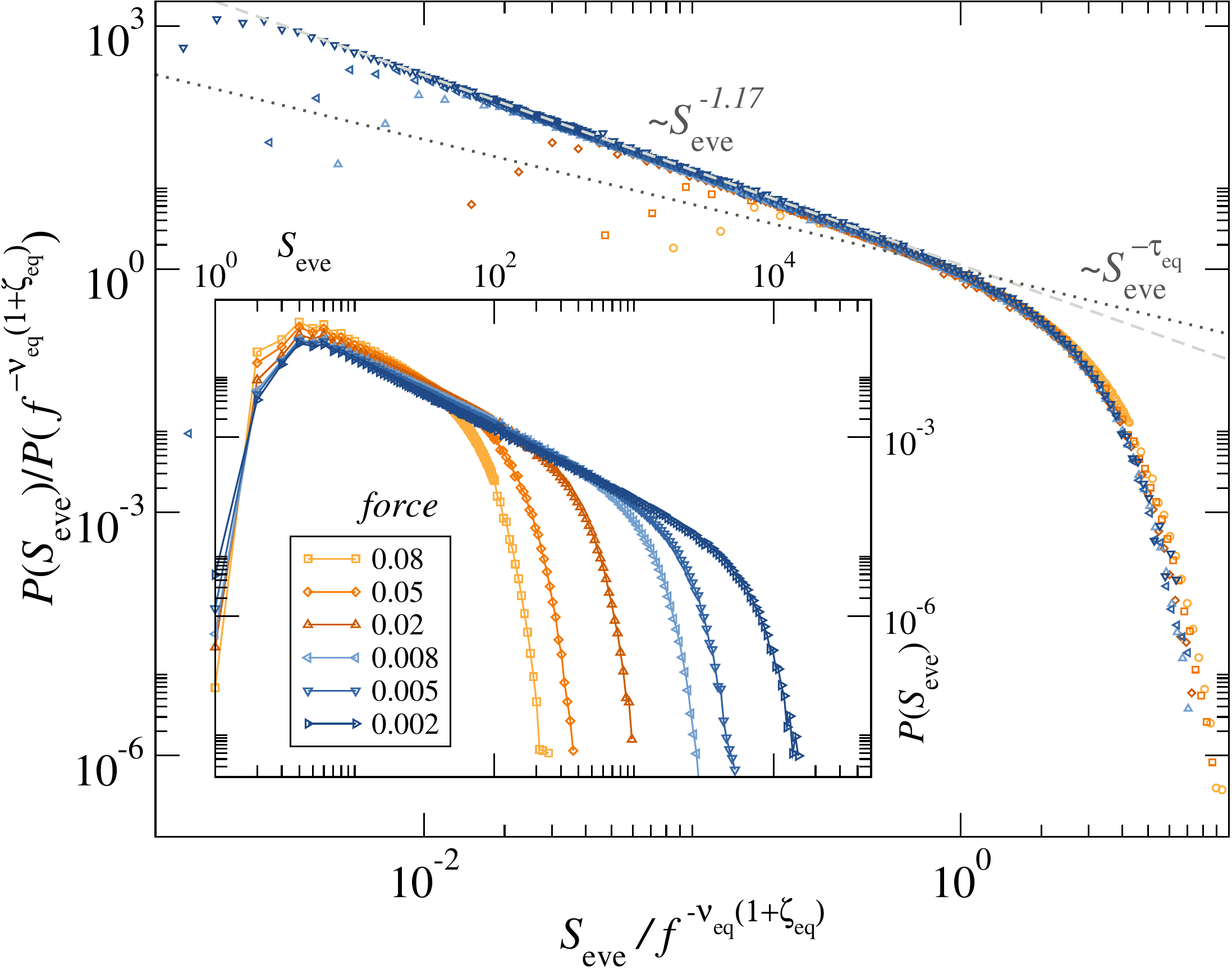}
  \caption{\label{fig:PofSeve}
  Events size distributions $P(\Seve)$ at different forces (inset) collapse by plotting
  $P(\Seve)/P(S_c)$ vs $\Seve/S_c$ with $S_c(f)=f^{-\nueq(1+\zetaeq)}$ (main panel),
  therefore validating the expected creep scaling $\Lopt \sim f^{-\nueq}$,
  given $S_c\sim \Lopt^{(1+\zetaeq)}$.
  }
 \end{figure}
\fi 
%
From the conventional picture of creep dynamics, one would expect that for small driving forces 
the event size fluctuates around a ``typical" value $\Lopt^{d+\zetaeq}$.
However, in Fig.\ref{fig:PofSeve}, we show that the event size distribution displays an unexpected
power law scaling:
\begin{equation}
\label{eq:PSeve}
 P(\Seve) \sim \Seve^{-\tau} G(\Seve/S_c)
\end{equation}
similar to the depinning one with a force dependent cut off $S_c(f)$. 
A good collapse of the distributions at different forces is found for $S_c(f) \sim f^{-\alpha}$ with $\alpha =1.25$. 
This scaling with force is perfectly consistent with the cutoff being $S_c \simeq \Sopt \sim \Lopt^{d+\zetaeq}$, 
that for $d=1$ yields $\alpha=(d+\zetaeq)\nueq=5/4$.
We conclude that at variance with  standard scaling, the characteristic lenght $\Lopt$
corresponds to the ``largest'', rather than the ``typical'' size of the irreversible events. 
However the creep law~(\ref{creepformula}) is not affected, since,
for activated dynamics, the velocity is controlled by the largest barriers and therefore by $\Lopt$.
 
A second important feature of $P(\Seve)$ is the power law decay.
A scaling argument, valid for elastic systems~\cite{LDMW09}, suggests that the cut-off exponent $\alpha$ and the power law
exponent $\tau$ should satisfy the relation $\tau=2-\frac{2\nueq}{\alpha}=2-2/(d+\zetaeq)$, in analogy to Eq.~(\ref{Eq_tau}).
Here the cut-off $\sim f^{-\alpha}$ is controlled by the distance to equilibrium $f=0$ and we would expect the value $\tau=\taueq=4/5$.
However this is not the case and we find a larger exponent $\tau=1.17\pm0.01$.
Such a distribution with $\tau>\taueq$, violating the scaling relation, expresses an excess of small events 
compared to what is expected \textit{a priori} for a distribution of fully independent avalanches.

To shed light on this issue, we further inspect Fig.~\ref{fig:Patterns}.
We observe that creep events are organized in compact spatio-temporal patterns in contrast 
with depinning avalanches
appear randomly distributed along the interface, as well illustrated by the \textit{activity maps}
that supplement the sequence of metastable configurations.
Furthermore, this correlations can be quantified, for example, by computing the mean squared displacement
of the epicenters of a sequence of events as a function of time sequence, as shown in the SM accompanying this Letter.

Remarkably, there is a similarity of such short-time correlations between events with
the ones observed in real earthquakes, where a large main shock is followed by a cascade
of aftershocks~\cite{Scholz_2002,AGGL_2016}.
In that case, presumably due to the presence of correlations, the sequence of events is not
homogeneous and the scaling relation between $\tau$ and $\alpha$ does not hold. 
In absence of correlations, the value of the Gutenberg-Richter exponent $b=\frac32(\tau-1)$
should be smaller than the mean field prediction $3/4$, but from seismic records one
gets~\cite{Scholz_2002,JLR_2014} $b\simeq 1$.
Accordingly, we find in our data that the effective creep exponent $\tau$ is
bigger than one predicted for avalanches at equilibrium.

\iffigures
 \begin{figure}
\centering
\includegraphics[width=0.93\columnwidth]{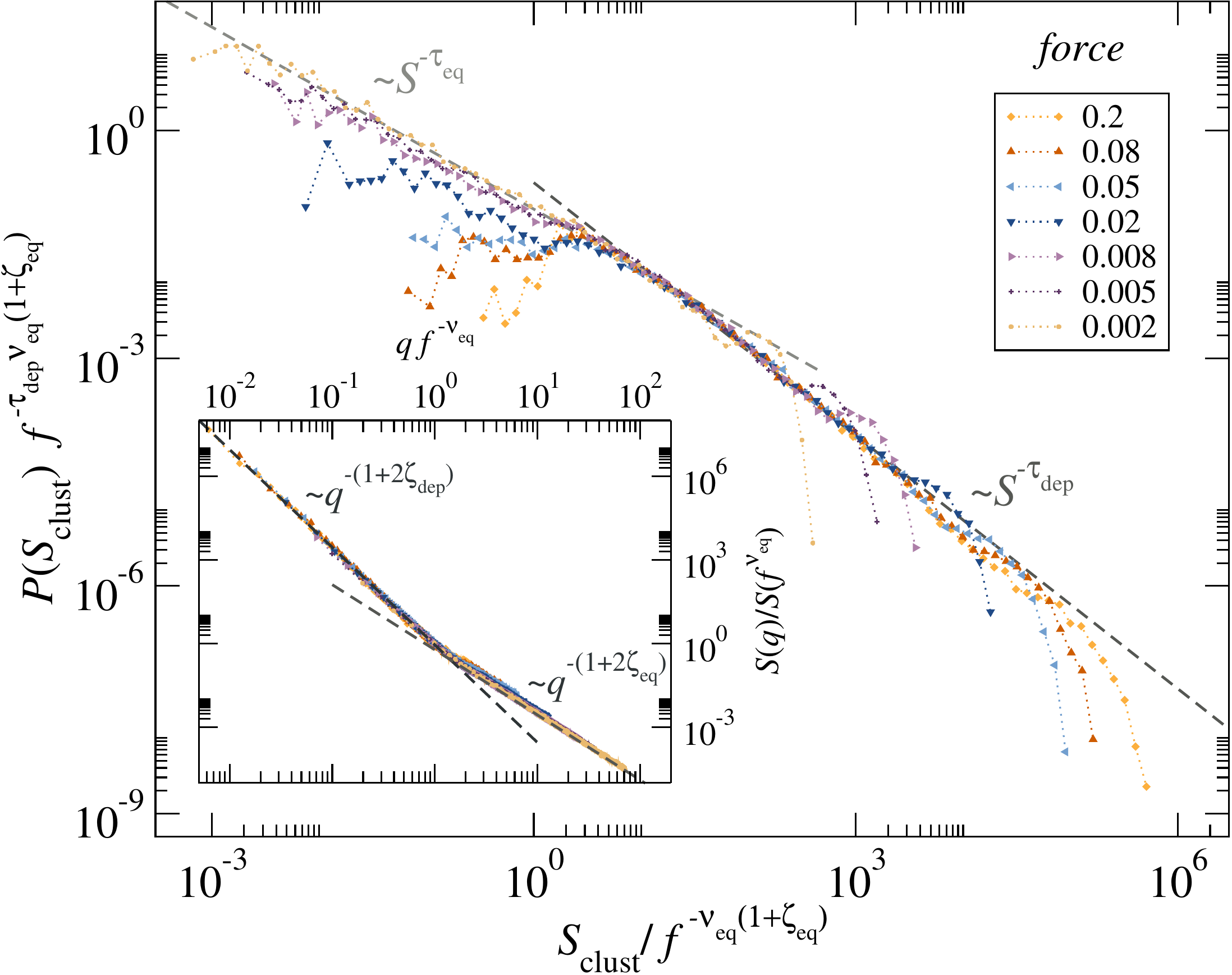}
  \caption{\label{fig:PofSclus} 
  Cluster area distribution $P(\Sclus)$ for different forces.
  A characteristic size $S_c\sim\Lopt^{(1+\zetaeq)}$, with
  $\Lopt\sim f^{-\nueq}$, separates small clusters that follow \textit{equilibrium}-like
  statistics from big clusters that follow a \textit{depinning}-like one.
  Inset: 
  The rescaled structure factor $S(q)$ for the same forces shows that $S(q)/S(l^{-1})$ is a function $q l$, 
  with $l=f^{-\nueq}\sim\Lopt$ denoting a geometrical crossover from \textit{equilibrium}-like
  roughness at small scales to a \textit{depinning}-like roughness at large scales. 
  }
 \end{figure}
\fi 

In order to analyze this spatio-temporal patterns we collect correlated events in \textit{clusters}
(creep events enclosed by a circle in Fig.~\ref{fig:Patterns}).
In presence of short range elasticity, a simple criterion for the cluster formation 
is to add a new event to a growing cluster if their spatial overlap is not null.
The growth of a cluster stops whenever a new event has no overlap with it.
This event, in turn, represents the seed for the creation of a new cluster.
In finite systems this procedure can generate system sized clusters, and in this case we decide
to break artificially the cluster construction, and start with a new cluster. 
This occurrence introduces a finite-size effect in the analysis, whose consequences, however,
are kept under control considering big systems.
Upon identifying the sequence of clusters, one can compute their size,
namely the sum of the areas of all the events that compose the cluster.

Figure~\ref{fig:PofSclus} shows the cluster size distribution, $P(\Sclus)$,
that displays a crossover at a value $\Sclus \sim \Sopt$.
Below $\Sopt$ we observe a power law with exponent $ \approx 0.80 \pm 0.06$
consistent with the equilibrium value, $\taueq=\frac{4}{5}$.
Above $\Sopt$, instead, the power-law exponent $\approx 1.11 \pm 0.04$ is
in good agreement with the depinning critical avalanches distribution
value~\cite{RLDW09} $\taudep \sim 1.11$.
In order to span more than eight decades in $\Sclus/\Sopt$, we have simulated a broad range of forces.
The equilibrium exponent appears in the limit of small forces, while forces larger than $f \simeq 0.05$
only display the depinning $\tau$ with a lower cutoff around $\Sopt$.
The upper cutoff of the distribution is controlled by the system size and
diverges in the thermodynamic limit,
as seen in a finite-size analysis~\cite{Ferrero_inprep}.

The inset of Fig.~\ref{fig:PofSclus} shows the structure factor ${S}(q)$ 
$S(q)=\overline{h_q h_{-q}}$ 
(here $h_q$ is the Fourier transform of a metastable configuration, $h(x)$, in the steady state 
and the overline stands for the average over many metastable configurations).
The figure clearly shows a crossover length scale $1/q_c \sim \Lopt$ that separates
short scales (large $q$), displaying an equilibrium roughness exponent $\zetaeq=2/3$,
from large scales (small $q$) displaying a depinning roughness $\zetadep \approx 1.25$, 
in agreement with Refs. \cite{Kardar1985} and \cite{Rosso-PRE2003,Ferrero2013}
respectively.
This geometrical crossover is compatible with the exponent crossover 
in the clusters size distribution and supports the idea of these objects
being depinning-like above a scale $\Lopt$.
%
The robustness of this conclusion is confirmed by the study of the Random Field
(RF) disorder case which belongs to a different universality class at equilibrium,
but share the same exponents as RB at the depinning transition~\cite{Chauve-PRB2000,NF93,Ferrero2013}
(see SM).

\smallskip \noindent {\it Discussion.} 
Our newly developed algorithm allows us to go deep in the creep regime of an elastic
interface moving in a disordered environment. It gives us an accurate description 
of the forward irreversible motion in terms of
a sequence of well defined activated events \textit{that goes far beyond the FGR analysis}.
The most striking property emerging from our study of creep events is
their occurrence in correlated spatio-temporal patterns,
in sharp contrast with depinning avalanches nucleating randomly along the line.
Despite the novel properties displayed by such dynamics, we find that the creep law
is verified.

We have constructed collective objects or ``clusters'' of creep events
and identified the sequence of these objects with the depinning-like 
motion predicted by the FRG analysis at a coarse-grain scale~\cite{Chauve-PRB2000}.
Due to the imposed limit of vanishing temperature, the cluster statistics
is scale-free in our study.
Still, for a finite temperature we expect a cutoff, related to the
temperature-controlled length scale predicted by the FRG.
Remarkably, we find out that this collective avalanches are
formed by correlated \textit{thermally activated} events and
not by a \textit{deterministic} gradient-descent motion.
Not being fully anticipated by FRG, our picture opens
a door for future theoretical studies.

We believe that the clustering behavior of creep events can be observed
in experiments with the current
magneto-optical techniques able to directly visualize the interface motion.
In fact, monitoring ion–irradiated magnetic thin films,
Repain \textit{et al.}~\cite{RBJFM04} were able to observe small correlated
events in the creep regime, their characteristic size increasing when
lowering the external field, in good qualitative agreement with our predictions.
Furthermore, we can quantitatively anticipate the observation of
spatially resolved individual creep events on Pt(0.35nm)/Co(0.45nm)/Pt(0.35nm)
films~\cite{GBFJKG14} at room temperature for domain wall velocities
of order $1nm/s$ (see SM).
We expect, not only direct visualizations, but also noise
measurements~\cite{Zapperi_Cizeau_1998, Zapperi_Durin_2000, DurinarXiv2016}
to allow for a full quantitative test of our predictions on creep dynamics. 

\smallskip \noindent {\it Acknowledgments.}
E.E.F acknowledge financial support from ERC Grant No. ADG20110209.
A.B.K. acknowledges partial support from Projects CONICET-PIP11220120100250CO
and ANPCyT-PICT-2011-1537 (Argentina).
This work was supported in part by the Swiss SNF under Division II. 
Most of the computations were performed using the Froggy platform of the CIMENT
infrastructure supported by the Rhône-Alpes region (Grant No. CPER07-13 CIRA)
and the Equip@Meso project (Reference No. ANR-10-EQPX-29-01).
We thank M.V. Duprez for her help on the schematic figures.
Further, we would like to warmly thank S. Bustingorry, J. Curiale, 
E.A. Jagla, V. Jeudy, V. Lecomte, and A. Mougin for fruitful discussions. 

\bibliography{DES}

\begin{thebibliography}{46}%
\makeatletter
\providecommand \@ifxundefined [1]{%
 \@ifx{#1\undefined}
}%
\providecommand \@ifnum [1]{%
 \ifnum #1\expandafter \@firstoftwo
 \else \expandafter \@secondoftwo
 \fi
}%
\providecommand \@ifx [1]{%
 \ifx #1\expandafter \@firstoftwo
 \else \expandafter \@secondoftwo
 \fi
}%
\providecommand \natexlab [1]{#1}%
\providecommand \enquote  [1]{``#1''}%
\providecommand \bibnamefont  [1]{#1}%
\providecommand \bibfnamefont [1]{#1}%
\providecommand \citenamefont [1]{#1}%
\providecommand \href@noop [0]{\@secondoftwo}%
\providecommand \href [0]{\begingroup \@sanitize@url \@href}%
\providecommand \@href[1]{\@@startlink{#1}\@@href}%
\providecommand \@@href[1]{\endgroup#1\@@endlink}%
\providecommand \@sanitize@url [0]{\catcode `\\12\catcode `\$12\catcode
  `\&12\catcode `\#12\catcode `\^12\catcode `\_12\catcode `\%12\relax}%
\providecommand \@@startlink[1]{}%
\providecommand \@@endlink[0]{}%
\providecommand \url  [0]{\begingroup\@sanitize@url \@url }%
\providecommand \@url [1]{\endgroup\@href {#1}{\urlprefix }}%
\providecommand \urlprefix  [0]{URL }%
\providecommand \Eprint [0]{\href }%
\providecommand \doibase [0]{http://dx.doi.org/}%
\providecommand \selectlanguage [0]{\@gobble}%
\providecommand \bibinfo  [0]{\@secondoftwo}%
\providecommand \bibfield  [0]{\@secondoftwo}%
\providecommand \translation [1]{[#1]}%
\providecommand \BibitemOpen [0]{}%
\providecommand \bibitemStop [0]{}%
\providecommand \bibitemNoStop [0]{.\EOS\space}%
\providecommand \EOS [0]{\spacefactor3000\relax}%
\providecommand \BibitemShut  [1]{\csname bibitem#1\endcsname}%
\let\auto@bib@innerbib\@empty
\bibitem [{\citenamefont {Lemerle}\ \emph {et~al.}(1998)\citenamefont
  {Lemerle}, \citenamefont {Ferr\'e}, \citenamefont {Chappert}, \citenamefont
  {Mathet}, \citenamefont {Giamarchi},\ and\ \citenamefont
  {Le~Doussal}}]{Lemerle-PRL1998}%
  \BibitemOpen
  \bibfield  {author} {\bibinfo {author} {\bibfnamefont {S.}~\bibnamefont
  {Lemerle}}, \bibinfo {author} {\bibfnamefont {J.}~\bibnamefont {Ferr\'e}},
  \bibinfo {author} {\bibfnamefont {C.}~\bibnamefont {Chappert}}, \bibinfo
  {author} {\bibfnamefont {V.}~\bibnamefont {Mathet}}, \bibinfo {author}
  {\bibfnamefont {T.}~\bibnamefont {Giamarchi}}, \ and\ \bibinfo {author}
  {\bibfnamefont {P.}~\bibnamefont {Le~Doussal}},\ }\href {\doibase
  10.1103/PhysRevLett.80.849} {\bibfield  {journal} {\bibinfo  {journal} {Phys.
  Rev. Lett.}\ }\textbf {\bibinfo {volume} {80}},\ \bibinfo {pages} {849}
  (\bibinfo {year} {1998})}\BibitemShut {NoStop}%
\bibitem [{\citenamefont {Yamanouchi}\ \emph {et~al.}(2006)\citenamefont
  {Yamanouchi}, \citenamefont {Chiba}, \citenamefont {Matsukura}, \citenamefont
  {Dietl},\ and\ \citenamefont {Ohno}}]{YCMDO06}%
  \BibitemOpen
  \bibfield  {author} {\bibinfo {author} {\bibfnamefont {M.}~\bibnamefont
  {Yamanouchi}}, \bibinfo {author} {\bibfnamefont {D.}~\bibnamefont {Chiba}},
  \bibinfo {author} {\bibfnamefont {F.}~\bibnamefont {Matsukura}}, \bibinfo
  {author} {\bibfnamefont {T.}~\bibnamefont {Dietl}}, \ and\ \bibinfo {author}
  {\bibfnamefont {H.}~\bibnamefont {Ohno}},\ }\href@noop {} {\bibfield
  {journal} {\bibinfo  {journal} {Physical review letters}\ }\textbf {\bibinfo
  {volume} {96}},\ \bibinfo {pages} {096601} (\bibinfo {year}
  {2006})}\BibitemShut {NoStop}%
\bibitem [{\citenamefont {Metaxas}\ \emph {et~al.}(2007)\citenamefont
  {Metaxas}, \citenamefont {Jamet}, \citenamefont {Mougin}, \citenamefont
  {Cormier}, \citenamefont {Ferr{\'e}}, \citenamefont {Baltz}, \citenamefont
  {Rodmacq}, \citenamefont {Dieny},\ and\ \citenamefont {Stamps}}]{MJMCF07}%
  \BibitemOpen
  \bibfield  {author} {\bibinfo {author} {\bibfnamefont {P.}~\bibnamefont
  {Metaxas}}, \bibinfo {author} {\bibfnamefont {J.}~\bibnamefont {Jamet}},
  \bibinfo {author} {\bibfnamefont {A.}~\bibnamefont {Mougin}}, \bibinfo
  {author} {\bibfnamefont {M.}~\bibnamefont {Cormier}}, \bibinfo {author}
  {\bibfnamefont {J.}~\bibnamefont {Ferr{\'e}}}, \bibinfo {author}
  {\bibfnamefont {V.}~\bibnamefont {Baltz}}, \bibinfo {author} {\bibfnamefont
  {B.}~\bibnamefont {Rodmacq}}, \bibinfo {author} {\bibfnamefont
  {B.}~\bibnamefont {Dieny}}, \ and\ \bibinfo {author} {\bibfnamefont
  {R.}~\bibnamefont {Stamps}},\ }\href@noop {} {\bibfield  {journal} {\bibinfo
  {journal} {Physical Review Letters}\ }\textbf {\bibinfo {volume} {99}},\
  \bibinfo {pages} {217208} (\bibinfo {year} {2007})}\BibitemShut {NoStop}%
\bibitem [{\citenamefont {Repain}\ \emph {et~al.}(2004)\citenamefont {Repain},
  \citenamefont {Bauer}, \citenamefont {Jamet}, \citenamefont {Ferr{\'e}},
  \citenamefont {Mougin}, \citenamefont {Chappert},\ and\ \citenamefont
  {Bernas}}]{RBJFM04}%
  \BibitemOpen
  \bibfield  {author} {\bibinfo {author} {\bibfnamefont {V.}~\bibnamefont
  {Repain}}, \bibinfo {author} {\bibfnamefont {M.}~\bibnamefont {Bauer}},
  \bibinfo {author} {\bibfnamefont {J.-P.}\ \bibnamefont {Jamet}}, \bibinfo
  {author} {\bibfnamefont {J.}~\bibnamefont {Ferr{\'e}}}, \bibinfo {author}
  {\bibfnamefont {A.}~\bibnamefont {Mougin}}, \bibinfo {author} {\bibfnamefont
  {C.}~\bibnamefont {Chappert}}, \ and\ \bibinfo {author} {\bibfnamefont
  {H.}~\bibnamefont {Bernas}},\ }\href@noop {} {\bibfield  {journal} {\bibinfo
  {journal} {EPL (Europhysics Letters)}\ }\textbf {\bibinfo {volume} {68}},\
  \bibinfo {pages} {460} (\bibinfo {year} {2004})}\BibitemShut {NoStop}%
\bibitem [{\citenamefont {Jeudy}\ \emph {et~al.}(2016)\citenamefont {Jeudy},
  \citenamefont {Mougin}, \citenamefont {Bustingorry}, \citenamefont
  {Savero~Torres}, \citenamefont {Gorchon}, \citenamefont {Kolton},
  \citenamefont {Lema\^{\i}tre},\ and\ \citenamefont {Jamet}}]{Jeudy2016}%
  \BibitemOpen
  \bibfield  {author} {\bibinfo {author} {\bibfnamefont {V.}~\bibnamefont
  {Jeudy}}, \bibinfo {author} {\bibfnamefont {A.}~\bibnamefont {Mougin}},
  \bibinfo {author} {\bibfnamefont {S.}~\bibnamefont {Bustingorry}}, \bibinfo
  {author} {\bibfnamefont {W.}~\bibnamefont {Savero~Torres}}, \bibinfo {author}
  {\bibfnamefont {J.}~\bibnamefont {Gorchon}}, \bibinfo {author} {\bibfnamefont
  {A.~B.}\ \bibnamefont {Kolton}}, \bibinfo {author} {\bibfnamefont
  {A.}~\bibnamefont {Lema\^{\i}tre}}, \ and\ \bibinfo {author} {\bibfnamefont
  {J.-P.}\ \bibnamefont {Jamet}},\ }\href {\doibase
  10.1103/PhysRevLett.117.057201} {\bibfield  {journal} {\bibinfo  {journal}
  {Phys. Rev. Lett.}\ }\textbf {\bibinfo {volume} {117}},\ \bibinfo {pages}
  {057201} (\bibinfo {year} {2016})}\BibitemShut {NoStop}%
\bibitem [{\citenamefont {Tybell}\ \emph {et~al.}(2002)\citenamefont {Tybell},
  \citenamefont {Paruch}, \citenamefont {Giamarchi},\ and\ \citenamefont
  {Triscone}}]{TPGT02}%
  \BibitemOpen
  \bibfield  {author} {\bibinfo {author} {\bibfnamefont {T.}~\bibnamefont
  {Tybell}}, \bibinfo {author} {\bibfnamefont {P.}~\bibnamefont {Paruch}},
  \bibinfo {author} {\bibfnamefont {T.}~\bibnamefont {Giamarchi}}, \ and\
  \bibinfo {author} {\bibfnamefont {J.-M.}\ \bibnamefont {Triscone}},\
  }\href@noop {} {\bibfield  {journal} {\bibinfo  {journal} {Phys. Rev. Lett.}\
  }\textbf {\bibinfo {volume} {89}},\ \bibinfo {pages} {097601} (\bibinfo
  {year} {2002})}\BibitemShut {NoStop}%
\bibitem [{\citenamefont {Paruch}\ and\ \citenamefont {Triscone}(2006)}]{PT06}%
  \BibitemOpen
  \bibfield  {author} {\bibinfo {author} {\bibfnamefont {P.}~\bibnamefont
  {Paruch}}\ and\ \bibinfo {author} {\bibfnamefont {J.-M.}\ \bibnamefont
  {Triscone}},\ }\href@noop {} {\bibfield  {journal} {\bibinfo  {journal}
  {Applied physics letters}\ }\textbf {\bibinfo {volume} {88}},\ \bibinfo
  {pages} {162907} (\bibinfo {year} {2006})}\BibitemShut {NoStop}%
\bibitem [{\citenamefont {Moulinet}\ \emph {et~al.}(2002)\citenamefont
  {Moulinet}, \citenamefont {Guthmann},\ and\ \citenamefont
  {Rolley}}]{MGR2002}%
  \BibitemOpen
  \bibfield  {author} {\bibinfo {author} {\bibfnamefont {S.}~\bibnamefont
  {Moulinet}}, \bibinfo {author} {\bibfnamefont {C.}~\bibnamefont {Guthmann}},
  \ and\ \bibinfo {author} {\bibfnamefont {E.}~\bibnamefont {Rolley}},\
  }\href@noop {} {\bibfield  {journal} {\bibinfo  {journal} {The European
  Physical Journal E}\ }\textbf {\bibinfo {volume} {8}},\ \bibinfo {pages}
  {437} (\bibinfo {year} {2002})}\BibitemShut {NoStop}%
\bibitem [{\citenamefont {Bouchaud}\ \emph {et~al.}(2002)\citenamefont
  {Bouchaud}, \citenamefont {Bouchaud}, \citenamefont {Fisher}, \citenamefont
  {Ramanathan},\ and\ \citenamefont {Rice}}]{BBFRR2002}%
  \BibitemOpen
  \bibfield  {author} {\bibinfo {author} {\bibfnamefont {E.}~\bibnamefont
  {Bouchaud}}, \bibinfo {author} {\bibfnamefont {J.}~\bibnamefont {Bouchaud}},
  \bibinfo {author} {\bibfnamefont {D.}~\bibnamefont {Fisher}}, \bibinfo
  {author} {\bibfnamefont {S.}~\bibnamefont {Ramanathan}}, \ and\ \bibinfo
  {author} {\bibfnamefont {J.}~\bibnamefont {Rice}},\ }\href@noop {} {\bibfield
   {journal} {\bibinfo  {journal} {Journal of the Mechanics and Physics of
  Solids}\ }\textbf {\bibinfo {volume} {50}},\ \bibinfo {pages} {1703}
  (\bibinfo {year} {2002})}\BibitemShut {NoStop}%
\bibitem [{\citenamefont {Zapperi}\ \emph {et~al.}(2006)\citenamefont
  {Zapperi}, \citenamefont {Alava},\ and\ \citenamefont {Nukala}}]{ZAN2006}%
  \BibitemOpen
  \bibfield  {author} {\bibinfo {author} {\bibfnamefont {S.}~\bibnamefont
  {Zapperi}}, \bibinfo {author} {\bibfnamefont {M.}~\bibnamefont {Alava}}, \
  and\ \bibinfo {author} {\bibfnamefont {P.}~\bibnamefont {Nukala}},\
  }\href@noop {} {\bibfield  {journal} {\bibinfo  {journal} {Advances in
  Physics}\ }\textbf {\bibinfo {volume} {55}},\ \bibinfo {pages} {349}
  (\bibinfo {year} {2006})}\BibitemShut {NoStop}%
\bibitem [{\citenamefont {Blatter}\ \emph {et~al.}(1994)\citenamefont
  {Blatter}, \citenamefont {Feigel'Man}, \citenamefont {Geshkenbein},
  \citenamefont {Larkin},\ and\ \citenamefont {Vinokur}}]{BFGLV94}%
  \BibitemOpen
  \bibfield  {author} {\bibinfo {author} {\bibfnamefont {G.}~\bibnamefont
  {Blatter}}, \bibinfo {author} {\bibfnamefont {M.}~\bibnamefont {Feigel'Man}},
  \bibinfo {author} {\bibfnamefont {V.}~\bibnamefont {Geshkenbein}}, \bibinfo
  {author} {\bibfnamefont {A.}~\bibnamefont {Larkin}}, \ and\ \bibinfo {author}
  {\bibfnamefont {V.~M.}\ \bibnamefont {Vinokur}},\ }\href@noop {} {\bibfield
  {journal} {\bibinfo  {journal} {Rev. Mod. Phys.}\ }\textbf {\bibinfo {volume}
  {66}},\ \bibinfo {pages} {1125} (\bibinfo {year} {1994})}\BibitemShut
  {NoStop}%
\bibitem [{\citenamefont {Parkin}\ \emph {et~al.}(2008)\citenamefont {Parkin},
  \citenamefont {Hayashi},\ and\ \citenamefont {Thomas}}]{Parkin-Science2008}%
  \BibitemOpen
  \bibfield  {author} {\bibinfo {author} {\bibfnamefont {S.~S.}\ \bibnamefont
  {Parkin}}, \bibinfo {author} {\bibfnamefont {M.}~\bibnamefont {Hayashi}}, \
  and\ \bibinfo {author} {\bibfnamefont {L.}~\bibnamefont {Thomas}},\ }\href
  {\doibase 10.1126/science.1145799} {\bibfield  {journal} {\bibinfo  {journal}
  {Science}\ }\textbf {\bibinfo {volume} {320}},\ \bibinfo {pages} {190}
  (\bibinfo {year} {2008})}\BibitemShut {NoStop}%
\bibitem [{\citenamefont {Ioffe}\ and\ \citenamefont
  {Vinokur}(1987)}]{Ioffe_Vinokur_1987}%
  \BibitemOpen
  \bibfield  {author} {\bibinfo {author} {\bibfnamefont {L.}~\bibnamefont
  {Ioffe}}\ and\ \bibinfo {author} {\bibfnamefont {V.}~\bibnamefont
  {Vinokur}},\ }\href@noop {} {\bibfield  {journal} {\bibinfo  {journal}
  {Journal of Physics C: Solid State Physics}\ }\textbf {\bibinfo {volume}
  {20}},\ \bibinfo {pages} {6149} (\bibinfo {year} {1987})}\BibitemShut
  {NoStop}%
\bibitem [{\citenamefont {Nattermann}(1987)}]{NattermannEPL1987}%
  \BibitemOpen
  \bibfield  {author} {\bibinfo {author} {\bibfnamefont {T.}~\bibnamefont
  {Nattermann}},\ }\href {http://stacks.iop.org/0295-5075/4/i=11/a=005}
  {\bibfield  {journal} {\bibinfo  {journal} {EPL (Europhysics Letters)}\
  }\textbf {\bibinfo {volume} {4}},\ \bibinfo {pages} {1241} (\bibinfo {year}
  {1987})}\BibitemShut {NoStop}%
\bibitem [{\citenamefont {Feigel'man}\ \emph {et~al.}(1989)\citenamefont
  {Feigel'man}, \citenamefont {Geshkenbein}, \citenamefont {Larkin},\ and\
  \citenamefont {Vinokur}}]{Feigelman1989}%
  \BibitemOpen
  \bibfield  {author} {\bibinfo {author} {\bibfnamefont {M.~V.}\ \bibnamefont
  {Feigel'man}}, \bibinfo {author} {\bibfnamefont {V.~B.}\ \bibnamefont
  {Geshkenbein}}, \bibinfo {author} {\bibfnamefont {A.~I.}\ \bibnamefont
  {Larkin}}, \ and\ \bibinfo {author} {\bibfnamefont {V.~M.}\ \bibnamefont
  {Vinokur}},\ }\href {\doibase 10.1103/PhysRevLett.63.2303} {\bibfield
  {journal} {\bibinfo  {journal} {Phys. Rev. Lett.}\ }\textbf {\bibinfo
  {volume} {63}},\ \bibinfo {pages} {2303} (\bibinfo {year}
  {1989})}\BibitemShut {NoStop}%
\bibitem [{\citenamefont {Nattermann}\ \emph
  {et~al.}(1990{\natexlab{a}})\citenamefont {Nattermann}, \citenamefont
  {Shapir},\ and\ \citenamefont {Vilfan}}]{NattermannPRB1990}%
  \BibitemOpen
  \bibfield  {author} {\bibinfo {author} {\bibfnamefont {T.}~\bibnamefont
  {Nattermann}}, \bibinfo {author} {\bibfnamefont {Y.}~\bibnamefont {Shapir}},
  \ and\ \bibinfo {author} {\bibfnamefont {I.}~\bibnamefont {Vilfan}},\ }\href
  {\doibase 10.1103/PhysRevB.42.8577} {\bibfield  {journal} {\bibinfo
  {journal} {Phys. Rev. B}\ }\textbf {\bibinfo {volume} {42}},\ \bibinfo
  {pages} {8577} (\bibinfo {year} {1990}{\natexlab{a}})}\BibitemShut {NoStop}%
\bibitem [{\citenamefont {Dong}\ \emph {et~al.}(1993)\citenamefont {Dong},
  \citenamefont {Marchetti}, \citenamefont {Middleton},\ and\ \citenamefont
  {Vinokur}}]{Dong1993}%
  \BibitemOpen
  \bibfield  {author} {\bibinfo {author} {\bibfnamefont {M.}~\bibnamefont
  {Dong}}, \bibinfo {author} {\bibfnamefont {M.}~\bibnamefont {Marchetti}},
  \bibinfo {author} {\bibfnamefont {A.~A.}\ \bibnamefont {Middleton}}, \ and\
  \bibinfo {author} {\bibfnamefont {V.}~\bibnamefont {Vinokur}},\ }\href@noop
  {} {\bibfield  {journal} {\bibinfo  {journal} {Physical review letters}\
  }\textbf {\bibinfo {volume} {70}},\ \bibinfo {pages} {662} (\bibinfo {year}
  {1993})}\BibitemShut {NoStop}%
\bibitem [{\citenamefont {Chauve}\ \emph {et~al.}(1998)\citenamefont {Chauve},
  \citenamefont {Giamarchi},\ and\ \citenamefont {Le~Doussal}}]{CGLD98}%
  \BibitemOpen
  \bibfield  {author} {\bibinfo {author} {\bibfnamefont {P.}~\bibnamefont
  {Chauve}}, \bibinfo {author} {\bibfnamefont {T.}~\bibnamefont {Giamarchi}}, \
  and\ \bibinfo {author} {\bibfnamefont {P.}~\bibnamefont {Le~Doussal}},\
  }\href@noop {} {\bibfield  {journal} {\bibinfo  {journal} {EPL (Europhysics
  Letters)}\ }\textbf {\bibinfo {volume} {44}},\ \bibinfo {pages} {110}
  (\bibinfo {year} {1998})}\BibitemShut {NoStop}%
\bibitem [{\citenamefont {Chauve}\ \emph {et~al.}(2000)\citenamefont {Chauve},
  \citenamefont {Giamarchi},\ and\ \citenamefont
  {Le~Doussal}}]{Chauve-PRB2000}%
  \BibitemOpen
  \bibfield  {author} {\bibinfo {author} {\bibfnamefont {P.}~\bibnamefont
  {Chauve}}, \bibinfo {author} {\bibfnamefont {T.}~\bibnamefont {Giamarchi}}, \
  and\ \bibinfo {author} {\bibfnamefont {P.}~\bibnamefont {Le~Doussal}},\
  }\href {\doibase 10.1103/PhysRevB.62.6241} {\bibfield  {journal} {\bibinfo
  {journal} {Phys. Rev. B}\ }\textbf {\bibinfo {volume} {62}},\ \bibinfo
  {pages} {6241} (\bibinfo {year} {2000})}\BibitemShut {NoStop}%
\bibitem [{\citenamefont {Scholz}(2002)}]{Scholz_2002}%
  \BibitemOpen
  \bibfield  {author} {\bibinfo {author} {\bibfnamefont {C.~H.}\ \bibnamefont
  {Scholz}},\ }\href@noop {} {\emph {\bibinfo {title} {The mechanics of
  earthquakes and faulting}}}\ (\bibinfo  {publisher} {Cambridge university
  press},\ \bibinfo {year} {2002})\BibitemShut {NoStop}%
\bibitem [{\citenamefont {Jagla}\ \emph {et~al.}(2014)\citenamefont {Jagla},
  \citenamefont {Landes},\ and\ \citenamefont {Rosso}}]{JLR_2014}%
  \BibitemOpen
  \bibfield  {author} {\bibinfo {author} {\bibfnamefont {E.}~\bibnamefont
  {Jagla}}, \bibinfo {author} {\bibfnamefont {F.~P.}\ \bibnamefont {Landes}}, \
  and\ \bibinfo {author} {\bibfnamefont {A.}~\bibnamefont {Rosso}},\
  }\href@noop {} {\bibfield  {journal} {\bibinfo  {journal} {Physical review
  letters}\ }\textbf {\bibinfo {volume} {112}},\ \bibinfo {pages} {174301}
  (\bibinfo {year} {2014})}\BibitemShut {NoStop}%
\bibitem [{\citenamefont {Jagla}\ and\ \citenamefont {Kolton}(2010)}]{JK_2010}%
  \BibitemOpen
  \bibfield  {author} {\bibinfo {author} {\bibfnamefont {E.~A.}\ \bibnamefont
  {Jagla}}\ and\ \bibinfo {author} {\bibfnamefont {A.~B.}\ \bibnamefont
  {Kolton}},\ }\href {\doibase 10.1029/2009JB006974} {\bibfield  {journal}
  {\bibinfo  {journal} {Journal of Geophysical Research: Solid Earth}\ }\textbf
  {\bibinfo {volume} {115}},\ \bibinfo {pages} {n/a} (\bibinfo {year}
  {2010})},\ \bibinfo {note} {b05312}\BibitemShut {NoStop}%
\bibitem [{\citenamefont {Gorchon}\ \emph {et~al.}(2014)\citenamefont
  {Gorchon}, \citenamefont {Bustingorry}, \citenamefont {Ferr{\'e}},
  \citenamefont {Jeudy}, \citenamefont {Kolton},\ and\ \citenamefont
  {Giamarchi}}]{GBFJKG14}%
  \BibitemOpen
  \bibfield  {author} {\bibinfo {author} {\bibfnamefont {J.}~\bibnamefont
  {Gorchon}}, \bibinfo {author} {\bibfnamefont {S.}~\bibnamefont
  {Bustingorry}}, \bibinfo {author} {\bibfnamefont {J.}~\bibnamefont
  {Ferr{\'e}}}, \bibinfo {author} {\bibfnamefont {V.}~\bibnamefont {Jeudy}},
  \bibinfo {author} {\bibfnamefont {A.}~\bibnamefont {Kolton}}, \ and\ \bibinfo
  {author} {\bibfnamefont {T.}~\bibnamefont {Giamarchi}},\ }\href@noop {}
  {\bibfield  {journal} {\bibinfo  {journal} {Phys. Rev. Lett.}\ }\textbf
  {\bibinfo {volume} {113}},\ \bibinfo {pages} {027205} (\bibinfo {year}
  {2014})}\BibitemShut {NoStop}%
\bibitem [{\citenamefont {Fisher}(1998)}]{Fisher_1998}%
  \BibitemOpen
  \bibfield  {author} {\bibinfo {author} {\bibfnamefont {D.~S.}\ \bibnamefont
  {Fisher}},\ }\href@noop {} {\bibfield  {journal} {\bibinfo  {journal}
  {Physics reports}\ }\textbf {\bibinfo {volume} {301}},\ \bibinfo {pages}
  {113} (\bibinfo {year} {1998})}\BibitemShut {NoStop}%
\bibitem [{\citenamefont {Kardar}(1999)}]{Kardar-PR1998}%
  \BibitemOpen
  \bibfield  {author} {\bibinfo {author} {\bibfnamefont {M.}~\bibnamefont
  {Kardar}},\ }\href@noop {} {\bibfield  {journal} {\bibinfo  {journal}
  {Physics Reports}\ }\textbf {\bibinfo {volume} {301}},\ \bibinfo {pages} {85}
  (\bibinfo {year} {1999})}\BibitemShut {NoStop}%
\bibitem [{\citenamefont {Ferrero}\ \emph
  {et~al.}(2013{\natexlab{a}})\citenamefont {Ferrero}, \citenamefont
  {Bustingorry}, \citenamefont {Kolton},\ and\ \citenamefont {Rosso}}]{FBKR13}%
  \BibitemOpen
  \bibfield  {author} {\bibinfo {author} {\bibfnamefont {E.~E.}\ \bibnamefont
  {Ferrero}}, \bibinfo {author} {\bibfnamefont {S.}~\bibnamefont
  {Bustingorry}}, \bibinfo {author} {\bibfnamefont {A.~B.}\ \bibnamefont
  {Kolton}}, \ and\ \bibinfo {author} {\bibfnamefont {A.}~\bibnamefont
  {Rosso}},\ }\href@noop {} {\bibfield  {journal} {\bibinfo  {journal} {Comptes
  Rendus Physique}\ }\textbf {\bibinfo {volume} {14}},\ \bibinfo {pages} {641}
  (\bibinfo {year} {2013}{\natexlab{a}})}\BibitemShut {NoStop}%
\bibitem [{\citenamefont {Narayan}\ and\ \citenamefont {Fisher}(1993)}]{NF93}%
  \BibitemOpen
  \bibfield  {author} {\bibinfo {author} {\bibfnamefont {O.}~\bibnamefont
  {Narayan}}\ and\ \bibinfo {author} {\bibfnamefont {D.~S.}\ \bibnamefont
  {Fisher}},\ }\href@noop {} {\bibfield  {journal} {\bibinfo  {journal}
  {Physical Review B}\ }\textbf {\bibinfo {volume} {48}},\ \bibinfo {pages}
  {7030} (\bibinfo {year} {1993})}\BibitemShut {NoStop}%
\bibitem [{\citenamefont {Rosso}\ \emph {et~al.}(2009)\citenamefont {Rosso},
  \citenamefont {Le~Doussal},\ and\ \citenamefont {Wiese}}]{RLDW09}%
  \BibitemOpen
  \bibfield  {author} {\bibinfo {author} {\bibfnamefont {A.}~\bibnamefont
  {Rosso}}, \bibinfo {author} {\bibfnamefont {P.}~\bibnamefont {Le~Doussal}}, \
  and\ \bibinfo {author} {\bibfnamefont {K.~J.}\ \bibnamefont {Wiese}},\
  }\href@noop {} {\bibfield  {journal} {\bibinfo  {journal} {Physical Review
  B}\ }\textbf {\bibinfo {volume} {80}},\ \bibinfo {pages} {144204} (\bibinfo
  {year} {2009})}\BibitemShut {NoStop}%
\bibitem [{\citenamefont {Kolton}\ \emph {et~al.}(2006)\citenamefont {Kolton},
  \citenamefont {Rosso}, \citenamefont {Giamarchi},\ and\ \citenamefont
  {Krauth}}]{Kolton-PRL2006}%
  \BibitemOpen
  \bibfield  {author} {\bibinfo {author} {\bibfnamefont {A.~B.}\ \bibnamefont
  {Kolton}}, \bibinfo {author} {\bibfnamefont {A.}~\bibnamefont {Rosso}},
  \bibinfo {author} {\bibfnamefont {T.}~\bibnamefont {Giamarchi}}, \ and\
  \bibinfo {author} {\bibfnamefont {W.}~\bibnamefont {Krauth}},\ }\href
  {\doibase 10.1103/PhysRevLett.97.057001} {\bibfield  {journal} {\bibinfo
  {journal} {Phys. Rev. Lett.}\ }\textbf {\bibinfo {volume} {97}},\ \bibinfo
  {pages} {057001} (\bibinfo {year} {2006})}\BibitemShut {NoStop}%
\bibitem [{\citenamefont {Kolton}\ \emph {et~al.}(2009)\citenamefont {Kolton},
  \citenamefont {Rosso}, \citenamefont {Giamarchi},\ and\ \citenamefont
  {Krauth}}]{Kolton-PRB2009}%
  \BibitemOpen
  \bibfield  {author} {\bibinfo {author} {\bibfnamefont {A.~B.}\ \bibnamefont
  {Kolton}}, \bibinfo {author} {\bibfnamefont {A.}~\bibnamefont {Rosso}},
  \bibinfo {author} {\bibfnamefont {T.}~\bibnamefont {Giamarchi}}, \ and\
  \bibinfo {author} {\bibfnamefont {W.}~\bibnamefont {Krauth}},\ }\href
  {\doibase 10.1103/PhysRevB.79.184207} {\bibfield  {journal} {\bibinfo
  {journal} {Phys. Rev. B}\ }\textbf {\bibinfo {volume} {79}},\ \bibinfo
  {pages} {184207} (\bibinfo {year} {2009})}\BibitemShut {NoStop}%
\bibitem [{\citenamefont {Le~Doussal}\ \emph {et~al.}(2009)\citenamefont
  {Le~Doussal}, \citenamefont {Middleton},\ and\ \citenamefont
  {Wiese}}]{LDMW09}%
  \BibitemOpen
  \bibfield  {author} {\bibinfo {author} {\bibfnamefont {P.}~\bibnamefont
  {Le~Doussal}}, \bibinfo {author} {\bibfnamefont {A.~A.}\ \bibnamefont
  {Middleton}}, \ and\ \bibinfo {author} {\bibfnamefont {K.~J.}\ \bibnamefont
  {Wiese}},\ }\href@noop {} {\bibfield  {journal} {\bibinfo  {journal}
  {Physical Review E}\ }\textbf {\bibinfo {volume} {79}},\ \bibinfo {pages}
  {050101} (\bibinfo {year} {2009})}\BibitemShut {NoStop}%
\bibitem [{\citenamefont {Arcangelis}\ \emph {et~al.}(2016)\citenamefont
  {Arcangelis}, \citenamefont {Godano}, \citenamefont {Grasso},\ and\
  \citenamefont {Lippiello}}]{AGGL_2016}%
  \BibitemOpen
  \bibfield  {author} {\bibinfo {author} {\bibfnamefont {L.}~\bibnamefont
  {Arcangelis}}, \bibinfo {author} {\bibfnamefont {C.}~\bibnamefont {Godano}},
  \bibinfo {author} {\bibfnamefont {J.~R.}\ \bibnamefont {Grasso}}, \ and\
  \bibinfo {author} {\bibfnamefont {E.}~\bibnamefont {Lippiello}},\ }\href@noop
  {} {\bibfield  {journal} {\bibinfo  {journal} {to be published in Physics
  Report}\ } (\bibinfo {year} {2016})}\BibitemShut {NoStop}%
\bibitem [{\citenamefont {Ferrero}\ \emph {et~al.}()\citenamefont {Ferrero},
  \citenamefont {Foini}, \citenamefont {Giamarchi}, \citenamefont {Kolton},\
  and\ \citenamefont {Rosso}}]{Ferrero_inprep}%
  \BibitemOpen
  \bibfield  {author} {\bibinfo {author} {\bibfnamefont {E.}~\bibnamefont
  {Ferrero}}, \bibinfo {author} {\bibfnamefont {L.}~\bibnamefont {Foini}},
  \bibinfo {author} {\bibfnamefont {T.}~\bibnamefont {Giamarchi}}, \bibinfo
  {author} {\bibfnamefont {A.}~\bibnamefont {Kolton}}, \ and\ \bibinfo {author}
  {\bibfnamefont {A.}~\bibnamefont {Rosso}},\ }\href@noop {} {\bibinfo
  {journal} {In preparation}\ }\BibitemShut {NoStop}%
\bibitem [{\citenamefont {Kardar}(1985)}]{Kardar1985}%
  \BibitemOpen
\bibfield  {journal} {  }\bibfield  {author} {\bibinfo {author} {\bibfnamefont
  {M.}~\bibnamefont {Kardar}},\ }\href {\doibase 10.1103/PhysRevLett.55.2923}
  {\bibfield  {journal} {\bibinfo  {journal} {Phys. Rev. Lett.}\ }\textbf
  {\bibinfo {volume} {55}},\ \bibinfo {pages} {2923} (\bibinfo {year}
  {1985})}\BibitemShut {NoStop}%
\bibitem [{\citenamefont {Rosso}\ \emph {et~al.}(2003)\citenamefont {Rosso},
  \citenamefont {Hartmann},\ and\ \citenamefont {Krauth}}]{Rosso-PRE2003}%
  \BibitemOpen
  \bibfield  {author} {\bibinfo {author} {\bibfnamefont {A.}~\bibnamefont
  {Rosso}}, \bibinfo {author} {\bibfnamefont {A.~K.}\ \bibnamefont {Hartmann}},
  \ and\ \bibinfo {author} {\bibfnamefont {W.}~\bibnamefont {Krauth}},\ }\href
  {\doibase 10.1103/PhysRevE.67.021602} {\bibfield  {journal} {\bibinfo
  {journal} {Phys. Rev. E}\ }\textbf {\bibinfo {volume} {67}},\ \bibinfo
  {pages} {021602} (\bibinfo {year} {2003})}\BibitemShut {NoStop}%
\bibitem [{\citenamefont {Ferrero}\ \emph
  {et~al.}(2013{\natexlab{b}})\citenamefont {Ferrero}, \citenamefont
  {Bustingorry},\ and\ \citenamefont {Kolton}}]{Ferrero2013}%
  \BibitemOpen
  \bibfield  {author} {\bibinfo {author} {\bibfnamefont {E.~E.}\ \bibnamefont
  {Ferrero}}, \bibinfo {author} {\bibfnamefont {S.}~\bibnamefont
  {Bustingorry}}, \ and\ \bibinfo {author} {\bibfnamefont {A.~B.}\ \bibnamefont
  {Kolton}},\ }\href {\doibase 10.1103/PhysRevE.87.032122} {\bibfield
  {journal} {\bibinfo  {journal} {Phys. Rev. E}\ }\textbf {\bibinfo {volume}
  {87}},\ \bibinfo {pages} {032122} (\bibinfo {year}
  {2013}{\natexlab{b}})}\BibitemShut {NoStop}%
\bibitem [{\citenamefont {Zapperi}\ \emph {et~al.}(1998)\citenamefont
  {Zapperi}, \citenamefont {Cizeau}, \citenamefont {Durin},\ and\ \citenamefont
  {Stanley}}]{Zapperi_Cizeau_1998}%
  \BibitemOpen
  \bibfield  {author} {\bibinfo {author} {\bibfnamefont {S.}~\bibnamefont
  {Zapperi}}, \bibinfo {author} {\bibfnamefont {P.}~\bibnamefont {Cizeau}},
  \bibinfo {author} {\bibfnamefont {G.}~\bibnamefont {Durin}}, \ and\ \bibinfo
  {author} {\bibfnamefont {H.~E.}\ \bibnamefont {Stanley}},\ }\href@noop {}
  {\bibfield  {journal} {\bibinfo  {journal} {Physical Review B}\ }\textbf
  {\bibinfo {volume} {58}},\ \bibinfo {pages} {6353} (\bibinfo {year}
  {1998})}\BibitemShut {NoStop}%
\bibitem [{\citenamefont {Durin}\ and\ \citenamefont
  {Zapperi}(2000)}]{Zapperi_Durin_2000}%
  \BibitemOpen
  \bibfield  {author} {\bibinfo {author} {\bibfnamefont {G.}~\bibnamefont
  {Durin}}\ and\ \bibinfo {author} {\bibfnamefont {S.}~\bibnamefont
  {Zapperi}},\ }\href@noop {} {\bibfield  {journal} {\bibinfo  {journal}
  {Physical Review Letters}\ }\textbf {\bibinfo {volume} {84}},\ \bibinfo
  {pages} {4705} (\bibinfo {year} {2000})}\BibitemShut {NoStop}%
\bibitem [{\citenamefont {{Durin}}\ \emph {et~al.}(2016)\citenamefont
  {{Durin}}, \citenamefont {{Bohn}}, \citenamefont {{Correa}}, \citenamefont
  {{Sommer}}, \citenamefont {{Le Doussal}},\ and\ \citenamefont
  {{Wiese}}}]{DurinarXiv2016}%
  \BibitemOpen
  \bibfield  {author} {\bibinfo {author} {\bibfnamefont {G.}~\bibnamefont
  {{Durin}}}, \bibinfo {author} {\bibfnamefont {F.}~\bibnamefont {{Bohn}}},
  \bibinfo {author} {\bibfnamefont {M.~A.}\ \bibnamefont {{Correa}}}, \bibinfo
  {author} {\bibfnamefont {R.~L.}\ \bibnamefont {{Sommer}}}, \bibinfo {author}
  {\bibfnamefont {P.}~\bibnamefont {{Le Doussal}}}, \ and\ \bibinfo {author}
  {\bibfnamefont {K.~J.}\ \bibnamefont {{Wiese}}},\ }\href@noop {} {\bibfield
  {journal} {\bibinfo  {journal} {ArXiv e-prints}\ } (\bibinfo {year}
  {2016})},\ \Eprint {http://arxiv.org/abs/1601.01331} {arXiv:1601.01331
  [cond-mat.mtrl-sci]} \BibitemShut {NoStop}%
\bibitem [{Note1()}]{Note1}%
  \BibitemOpen
  \bibinfo {note} {This constraint also induces an harmonic-anharmonic
  depinning crossover at large enough scales, that we consider together with
  finite size effects in~\cite {Ferrero_inprep}}\BibitemShut {NoStop}%
\bibitem [{\citenamefont {Fisher}(1986)}]{F86}%
  \BibitemOpen
  \bibfield  {author} {\bibinfo {author} {\bibfnamefont {D.~S.}\ \bibnamefont
  {Fisher}},\ }\href@noop {} {\bibfield  {journal} {\bibinfo  {journal} {Phys.
  Rev. Lett.}\ }\textbf {\bibinfo {volume} {56}},\ \bibinfo {pages} {1964}
  (\bibinfo {year} {1986})}\BibitemShut {NoStop}%
\bibitem [{\citenamefont {Chauve}\ \emph {et~al.}(2001)\citenamefont {Chauve},
  \citenamefont {Le~Doussal},\ and\ \citenamefont {J\"org~Wiese}}]{Chauve2001}%
  \BibitemOpen
  \bibfield  {author} {\bibinfo {author} {\bibfnamefont {P.}~\bibnamefont
  {Chauve}}, \bibinfo {author} {\bibfnamefont {P.}~\bibnamefont {Le~Doussal}},
  \ and\ \bibinfo {author} {\bibfnamefont {K.}~\bibnamefont {J\"org~Wiese}},\
  }\href {\doibase 10.1103/PhysRevLett.86.1785} {\bibfield  {journal} {\bibinfo
   {journal} {Phys. Rev. Lett.}\ }\textbf {\bibinfo {volume} {86}},\ \bibinfo
  {pages} {1785} (\bibinfo {year} {2001})}\BibitemShut {NoStop}%
\bibitem [{\citenamefont {Larkin}\ and\ \citenamefont
  {Ovchinnikov}(1979)}]{larkin1979pinning}%
  \BibitemOpen
  \bibfield  {author} {\bibinfo {author} {\bibfnamefont {A.}~\bibnamefont
  {Larkin}}\ and\ \bibinfo {author} {\bibfnamefont {Y.~N.}\ \bibnamefont
  {Ovchinnikov}},\ }\href@noop {} {\bibfield  {journal} {\bibinfo  {journal}
  {Journal of Low Temperature Physics}\ }\textbf {\bibinfo {volume} {34}},\
  \bibinfo {pages} {409} (\bibinfo {year} {1979})}\BibitemShut {NoStop}%
\bibitem [{\citenamefont {Nattermann}\ \emph
  {et~al.}(1990{\natexlab{b}})\citenamefont {Nattermann}, \citenamefont
  {Shapir},\ and\ \citenamefont {Vilfan}}]{nattermann1990interface}%
  \BibitemOpen
  \bibfield  {author} {\bibinfo {author} {\bibfnamefont {T.}~\bibnamefont
  {Nattermann}}, \bibinfo {author} {\bibfnamefont {Y.}~\bibnamefont {Shapir}},
  \ and\ \bibinfo {author} {\bibfnamefont {I.}~\bibnamefont {Vilfan}},\
  }\href@noop {} {\bibfield  {journal} {\bibinfo  {journal} {Physical Review
  B}\ }\textbf {\bibinfo {volume} {42}},\ \bibinfo {pages} {8577} (\bibinfo
  {year} {1990}{\natexlab{b}})}\BibitemShut {NoStop}%
\bibitem [{\citenamefont {D{\'e}mery}\ \emph {et~al.}(2014)\citenamefont
  {D{\'e}mery}, \citenamefont {Rosso},\ and\ \citenamefont
  {Ponson}}]{demery2014microstructural}%
  \BibitemOpen
  \bibfield  {author} {\bibinfo {author} {\bibfnamefont {V.}~\bibnamefont
  {D{\'e}mery}}, \bibinfo {author} {\bibfnamefont {A.}~\bibnamefont {Rosso}}, \
  and\ \bibinfo {author} {\bibfnamefont {L.}~\bibnamefont {Ponson}},\
  }\href@noop {} {\bibfield  {journal} {\bibinfo  {journal} {EPL (Europhysics
  Letters)}\ }\textbf {\bibinfo {volume} {105}},\ \bibinfo {pages} {34003}
  (\bibinfo {year} {2014})}\BibitemShut {NoStop}%
\bibitem [{\citenamefont {Kim}\ \emph {et~al.}(2009)\citenamefont {Kim},
  \citenamefont {Lee}, \citenamefont {Ahn}, \citenamefont {Lee}, \citenamefont
  {Lee}, \citenamefont {Cho}, \citenamefont {Seo}, \citenamefont {Shin},
  \citenamefont {Choe},\ and\ \citenamefont {Lee}}]{KimNature2009}%
  \BibitemOpen
  \bibfield  {author} {\bibinfo {author} {\bibfnamefont {K.-J.}\ \bibnamefont
  {Kim}}, \bibinfo {author} {\bibfnamefont {J.-C.}\ \bibnamefont {Lee}},
  \bibinfo {author} {\bibfnamefont {S.-M.}\ \bibnamefont {Ahn}}, \bibinfo
  {author} {\bibfnamefont {K.-S.}\ \bibnamefont {Lee}}, \bibinfo {author}
  {\bibfnamefont {C.-W.}\ \bibnamefont {Lee}}, \bibinfo {author} {\bibfnamefont
  {Y.~J.}\ \bibnamefont {Cho}}, \bibinfo {author} {\bibfnamefont
  {S.}~\bibnamefont {Seo}}, \bibinfo {author} {\bibfnamefont {K.-H.}\
  \bibnamefont {Shin}}, \bibinfo {author} {\bibfnamefont {S.-B.}\ \bibnamefont
  {Choe}}, \ and\ \bibinfo {author} {\bibfnamefont {H.-W.}\ \bibnamefont
  {Lee}},\ }\href@noop {} {\bibfield  {journal} {\bibinfo  {journal} {Nature}\
  }\textbf {\bibinfo {volume} {458}},\ \bibinfo {pages} {740} (\bibinfo {year}
  {2009})}\BibitemShut {NoStop}%
\end{thebibliography}%

\clearpage
\onecolumngrid
\begin{center}
\textbf{\large Supporting Information for ``Spatio-temporal patterns in ultra-slow domain wall creep dynamics''}\\[0.5cm]
\end{center}

\setcounter{equation}{0}
\setcounter{figure}{0}
\setcounter{table}{0}
\setcounter{section}{0}
\setcounter{page}{1}
\renewcommand{\thefigure}{S\arabic{figure}}
\renewcommand{\thetable}{S\arabic{table}}
\renewcommand{\theequation}{S\arabic{equation}}
\renewcommand{\thesection}{\Alph{section}}

\begin{center}
Ezequiel E. Ferrero, Laura Foini, Thierry Giamarchi, Alejandro B. Kolton, Alberto Rosso
\end{center}

\section*{\small abstract}
\begin{center}%
\begin{minipage}{0.8\textwidth}
 In this supplement we describe in details and validate
our model, we characterize the spatio-temporal correlations
between individual creep events and confirm that creep
motion is described by both depinning and equilibrium
exponents regardless of the RB or RF nature of the disorder.
Estimates of the relevant scales for the experimental 
test of our results in Pt/Co/Pt thin ferromagnetic films 
are provided.
\end{minipage}
\end{center}
\vspace{0.5cm}

\twocolumngrid

\section{Model details}

\iffigures
 \begin{figure}
 \center
 \includegraphics[width=0.8\columnwidth]{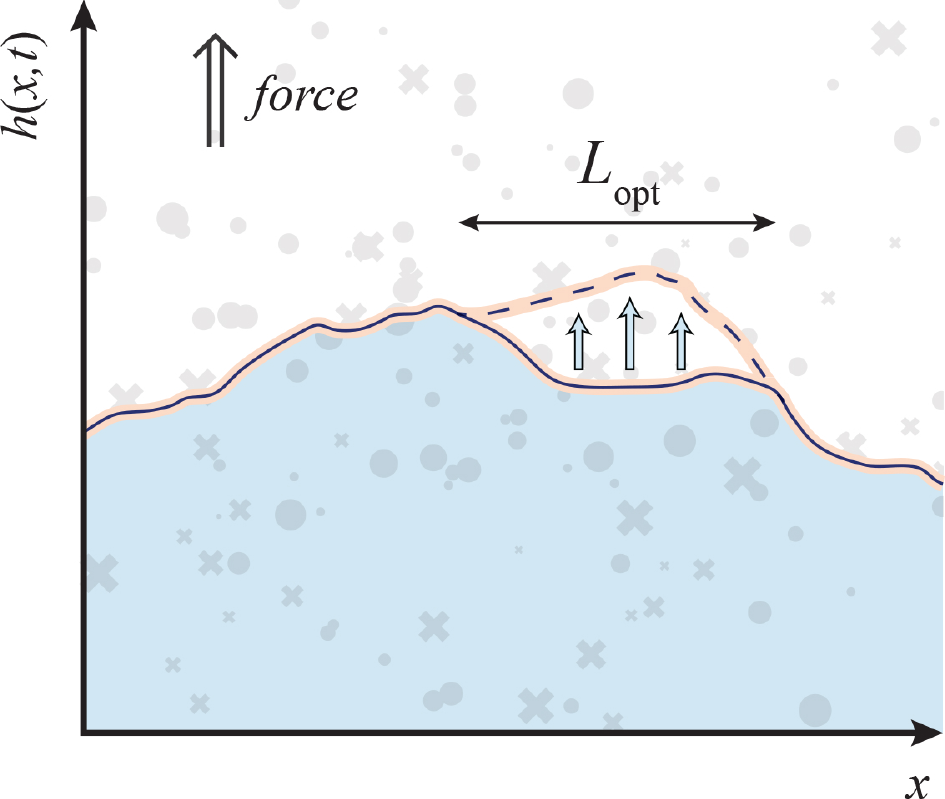} 
   \caption{\label{fig:LineSchem}
   \textit{Driven elastic line scheme~--}
   Sketch of an elastic line pinned at time $t$ in a metastable configuration $h(x,t)$
   and pushed homogeneously by an external force $f$.
   Circles and crosses in the background illustrate the quenched disorder.
   Below $\fc$ forward motion is possible thanks to thermal activation.
   In this picture, by a rearrangement of linear size $\sim\Lopt$ the line
   jumps in a new metastable state (dashed line) characterized by a smaller energy.
   }
 \end{figure}
\fi

The interface schematically depicted in  Fig.\ref{fig:LineSchem} 
is modeled as a discrete polymer of $L$ monomers 
at integer positions $h(i)$ ($i=0, \ldots, L-1$).
The string energy is given by:
\begin{equation}
E = \sum_i \left[ (h(i+1)- h(i))^2  - f h(i) + V(i,h(i)) \right].
\label{eq:energy}
\end{equation}
 
We consider periodic boundary conditions in the longitudinal direction
($h(L) \equiv h(0)$), and implement a hard metric
constraint ($|h(i)-h(i-1)| \le 1$) which significantly reduces the
configuration space~\footnote{
This constraint also induces an harmonic-anharmonic	depinning crossover
at large enough scales, that we consider together with finite size effects
in~\cite{Ferrero_inprep}
}.
The disorder $V(h(i), i)$ is computed from uncorrelated 
Gaussian numbers $R_{j,i}$ with zero mean and unit variance. 
To model RB disorder we defined $V_{RB}(h, i)=R_{h,i}$,
while for RF disorder we define 
$V_{RF}(h, i)=\sum_{k=0}^{h} R_{k,i}$,
such that $\overline{[V_{RF}(j, i)-V_{RF}(j', i')]^2}=\delta_{i,i'} |j-j'|$.

The two-steps polymer position update is performed as follows:

{\it (i) Activation}:
Starting from any metastable state we find the smallest compact
rearrangement that decreases the energy.
In order to do that, we fix a window $w$ and compute the optimal
path between two generic points $i,i+w$ of the polymer using
the Dijktra's algorithm adapted to compute the minimal energy
polymer between two fixed points~\cite{Kolton-PRB2009} (we do it for all $i$).
The minimal favorable rearrangement corresponds to the minimal window
for which the best path differs from the polymer configuration.
Within this approach, as discussed in the main article, we assume that
the minimal energy barrier identifies with the smallest compact
movement that decreases the energy.

{\it (ii) Deterministic relaxation}:
After the above activated move, the string is not necessarily in a new
metastable state;
so we let the line relax deterministically with a protocol
of elementary moves~\cite{Rosso-PRE2003}, the same dynamics
as implemented in~\cite{Kolton-PRB2009}.

The approximation introduced in the activated step allowed
us to overcome the severe computation limitations
of the exact algorithm and made it possible not only 
to increase of a factor $30$ the system size, but, and more importantly, 
to decrease of a factor $100$ the external drive $f$, unveiling the statistics and the clustering 
of the  activated events. 
Indeed, as discussed in~\cite{Kolton-PRL2006,Kolton-PRB2009}, 
the computational cost of the activated step  grows exponentially 
in $\Lopt(f)$ with the exact algorithm while, within our approximation, it has 
a polynomial cost.
Of course, there is a price to pay:
We loose access to the actual energy barrier values and
therefore to the real time.

On top of the algorithmic improvement, the ensemble of our code is
implemented to run in parallel in general-purpose Graphics
Processing Units, what gives an extra speedup to the simulations.
Source files are freely available under request.

\iffigures
\begin{figure*}[ht]
  \centering
   \includegraphics[width=0.42\textwidth]{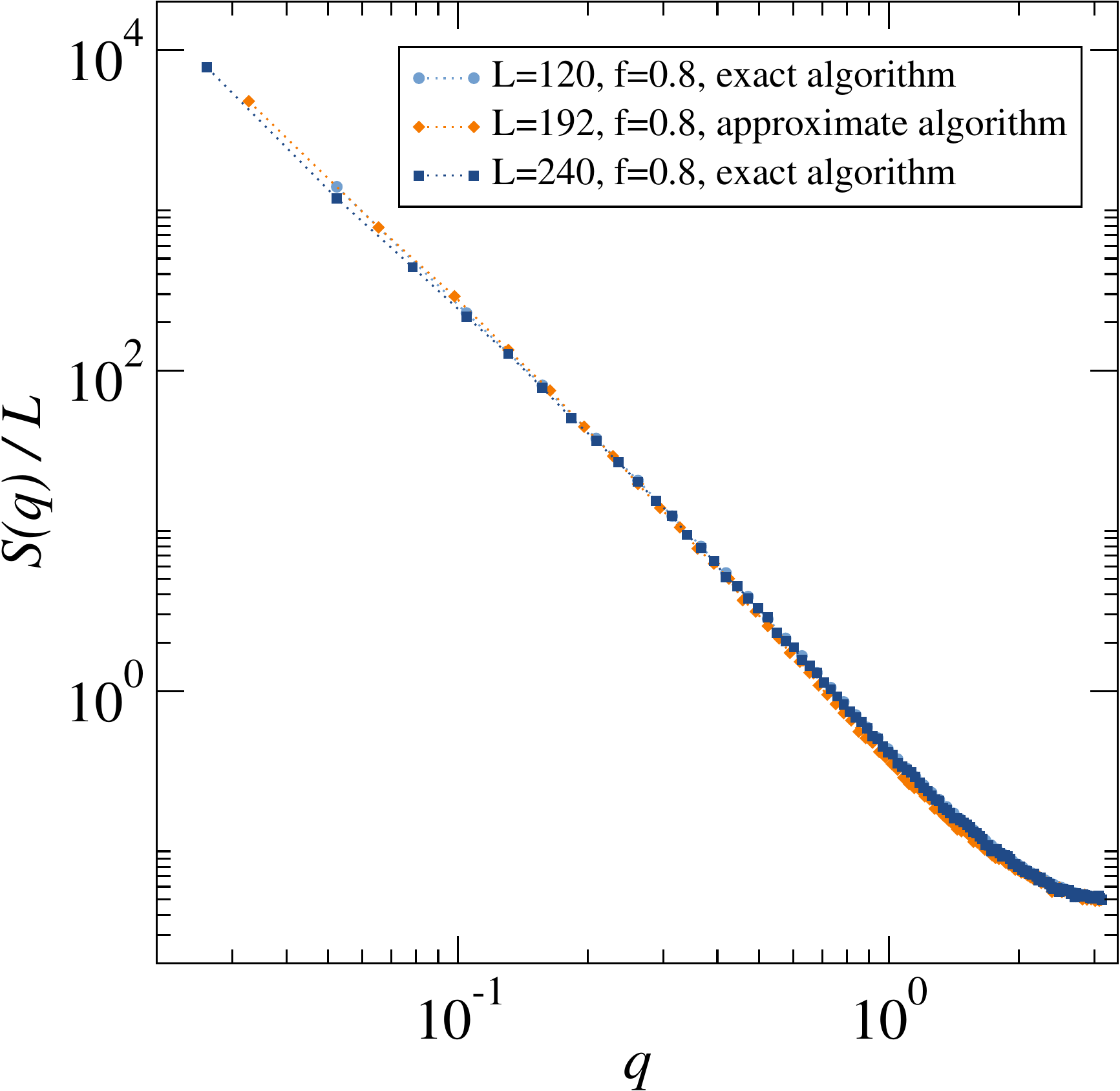}
      \hspace{0.2cm}
   \includegraphics[width=0.42\textwidth]{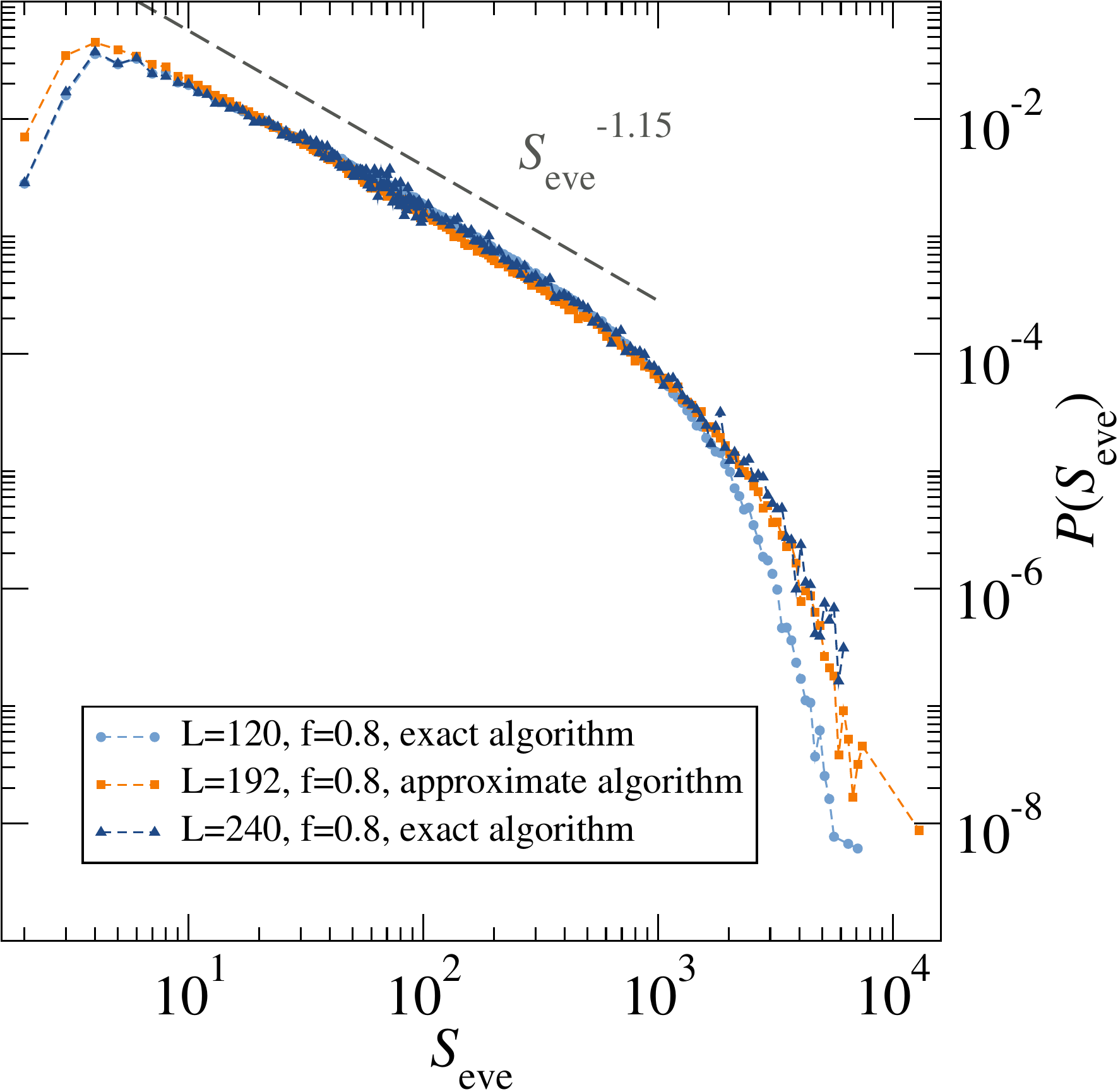}
  \caption{\label{fig:exact_vs_approximate}
  \textit{Left:} Average structure factors in the steady-state creep motion at $f=0.8$. 
  We compare the structure factor obtained with the approximate algorithm 
  for $L=192$, with the ones obtained with the exact algorithm for $L=120,240$.
  \textit{Right:} Comparison of the creep event size distribution obtained with the exact
  and approximate algorithms for a driving force $f=0.8$.
  The power law $P(\Seve) \sim \Seve^{-1.15}$ is shown as a guide to the eye.
  }
 \end{figure*}
\fi

To validate our approach we contrast it directly with the exact transition pathways algorithm
used in Refs.~\cite{Kolton-PRL2006,Kolton-PRB2009}.
In Fig.\ref{fig:exact_vs_approximate}-left and \ref{fig:exact_vs_approximate}-right 
we compare, respectively, the average structure factor and the event size distribution
at a given force.
A statistical difference can only be appreciated for small events
(small length scales and large wave-vectors $q$), presumably because of disregarding
the existence of large energy barriers for some of the small rearrangements.
In general the equivalence between the scaling of rearrangement sizes and energy barriers seems
to work very well all across the range of parameters where it is possible to simulate both algorithms.
Most importantly, differences at large scales, the ones that dominate the universal behavior
we aim to study, are unobservable.

\section{Temporal correlations between creep events}
\iffigures
\begin{figure}[hb]
\centering
\includegraphics[width=0.8\columnwidth]{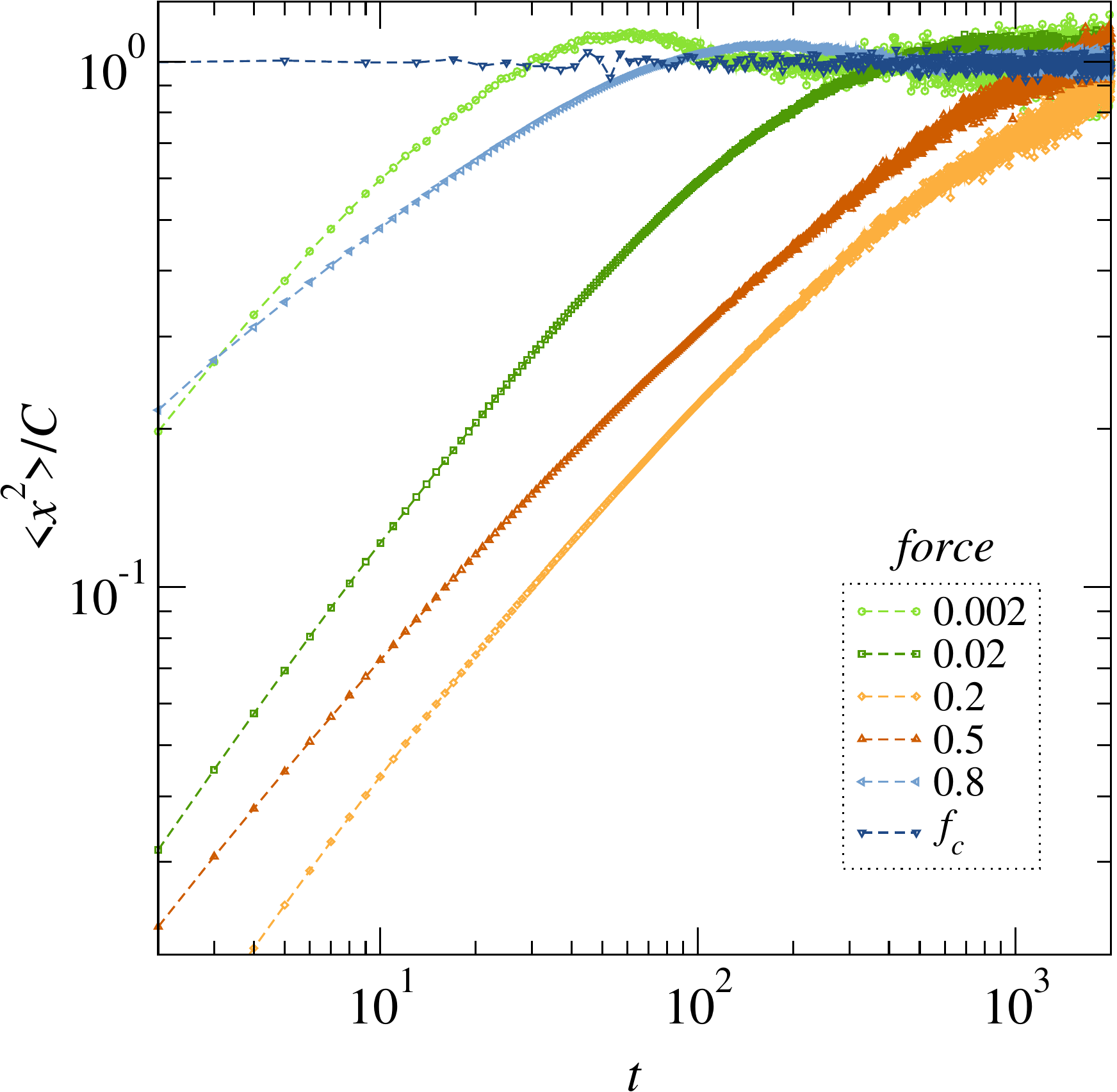}
  \caption{\label{fig:MSD}
Mean square displacement (MSD) $\left<x^2\right>$ of the events epicenter as a function of the observed window $t$
  (a discrete time distance in the ordered events sequence) for different forces.
  Correlations among consecutive events render the MSD non trivial in a force dependent time window.
  Fully uncorrelated events (as the ones for $f=f_c$) can only build a flat MSD.
  Data correspond to a system size $L=960$.
  }
\end{figure}
\fi 

In the activity maps shown in Figure \ref{fig:Patterns}
of the main article we observed that consecutive activated creep
events tend to be close in space, in sharp contrast with the 
behavior of consecutive depinning avalanches.
In order to make this qualitative comparison more quantitative
we analyze here the degree of temporal correlation among creep events
by computing the mean squared displacement (MSD),
$\langle \Delta X^2(t) \rangle$, of the (epicenter) location of events
separated by $t$ metastable states.
In absence of correlations we expect $\langle \Delta X^2(t) \rangle = C$
with $C = (1/L) \sum_{i=1}^{L/2} 2 i^2 = (L+1)(L+2)/12$,
for a line of size $L$.
As we show in Fig.~\ref{fig:MSD} this is well verified by depinning
avalanches at all ``times'', while creep events show this behavior
only asymptotically for large $t$.
In fact, at short times, the distance among epicenters tends to be small
as the events overlap in space,
reducing in this way the MSD compared to the uncorrelated case. 
Furthermore, one sees that for a fixed short time, the MSD displays
a non-monotonic behavior as a function of $f$.
This can be understood by considering that for consecutive events
overlapping in space the mean distance among their epicenters is
controlled by their typical size $\Leve(f)$, which has also a
non-monotonic behavior with $f$
(see Fig.~\ref{fig:cutoff} and the discussion here below).

\section{Random bond and random field disorder}\label{Section_RF}

Being an intermediate regime between two fixed points of the dynamics~\cite{CGLD98},
the creep is thus described by both equilibrium and depinning
exponents, rather than by different ones.

\begin{table}[ht]
\begin{small}
\hspace{0.3cm}
\begin{center}
 \begin{tabular}{lcc}
	\hline
	\hline
	\textbf{1d RB exponents}&\textbf{estimate}\\ 
	\hline
	$\zetaeq$ & $2/3$  \\
	$\nueq = 1/(2-\zetaeq)$ & $3/4$   \\ 
	$\taueq=2-2/(1+\zetaeq)$ & $4/5$  \\
	$\thetaeq=2\zetaeq-1$ & $1/3$  \\
	$\mu=\thetaeq \nueq$ & $1/4$  \\
	\hline
	\hline
	\textbf{1d RF exponents} & \textbf{estimate}   \\
	\hline
	$\zetaeqRF$ & $1$  \\
	$\nueqRF = 1/(2-\zetaeqRF)$ & $1$  \\ 
	$\taueqRF=2-2/(1+\zetaeqRF)$ & $1$  \\
	$\thetaeqRF=2\zetaeqRF-1$ & $1$  \\
	$\muRF=\thetaeqRF \nueqRF$ & $1$  \\
 	\hline
	\hline
	\textbf{1d RB \& RF exponents} & \textbf{estimate}   \\
	\hline
	$\zetadep$   & $1.250$ \\	
	$\nudep=1/(2-\zetadep)$ & $1.333$ \\
	$\taudep=2-2/(1+\zetadep)$ & $1.11$  \\
 	\hline
 \end{tabular}
\caption{
\label{tab:exponents}
Universal exponents relevant for the one dimensional creep motion, 
for short range elastic interactions, according to the disorder type. 
} 
\end{center}
 \end{small}
\end{table}

In Table~\ref{tab:exponents} we report the equilibrium and depinning
exponents~\cite{Kardar1985,F86,NF93,Chauve2001,Rosso-PRE2003,Kolton-PRL2006,Kolton-PRB2009,Ferrero2013}
for RB and RF.
It is worth remarking that RB and RF share 
the same depinning universality class while 
at equilibrium they display different exponents.

\iffigures
\begin{figure}
  \centering
 \includegraphics[width=0.8\columnwidth]{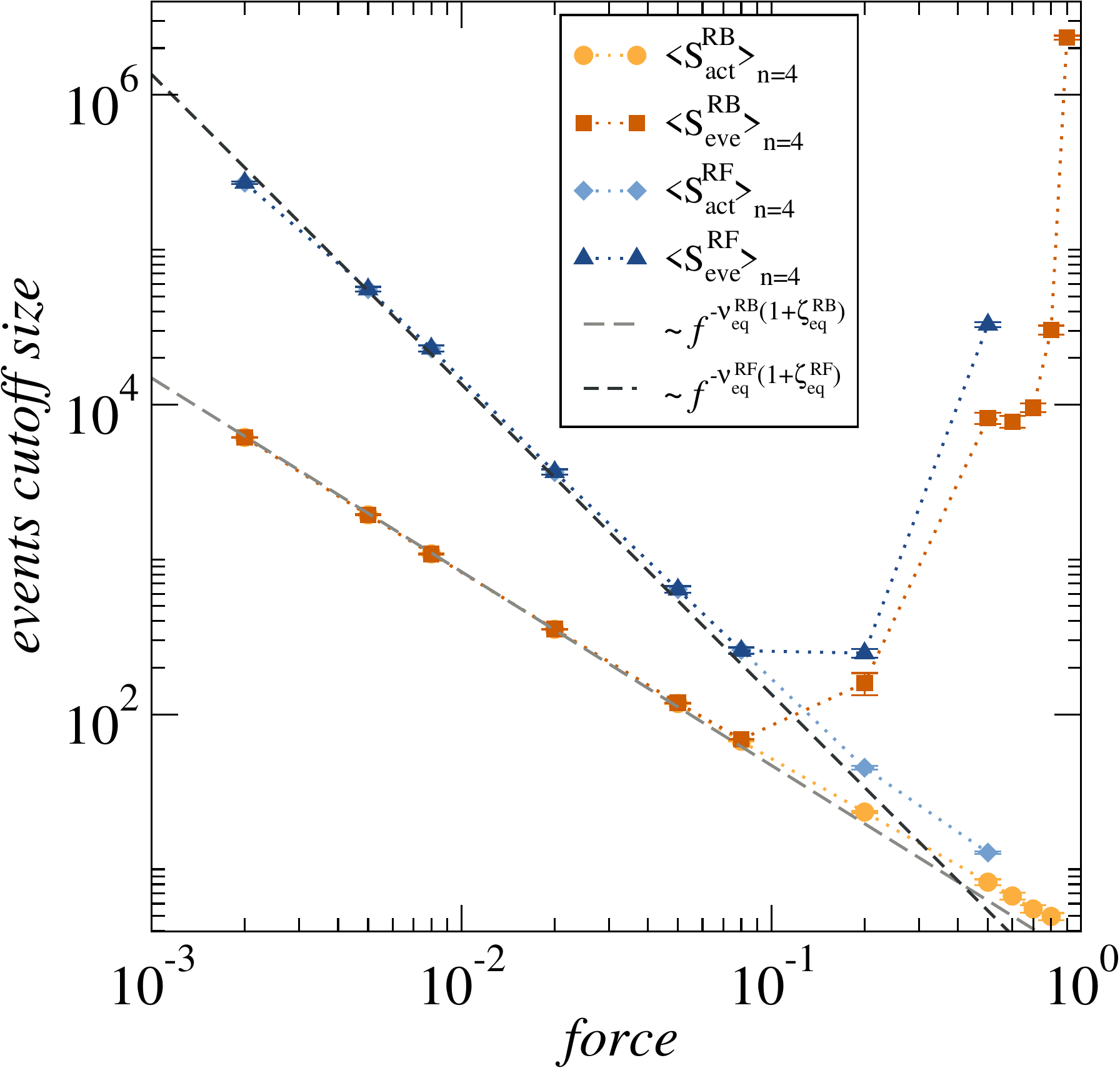}
  \caption{\label{fig:cutoff}
  Cutoff of the distribution of the areas of creep events for RB and RF disorders.
  In each case the mean value of fourth order $\left< X \right>_{n=4}=\left< X^4 \right>/\left< X^3 \right>$ is shown.
  The error bars are estimated from the difference with the values given by $\left< X \right>_{n=3}$.
  Dashed lines show the universal power-law divergence expected as $f \to 0$ from the phenomenological creep theory.
  Note that $\Seve \approx \Sact \sim \Sopt = \Lopt^{1+\zeta}$ at low forces, while at higher values $\Seve \gg \Sopt$ due to
  large deterministic relaxation.
  In particular one observes $\Sact\sim\Sopt \to \infty$ for $f\to 0$ and $\Seve\to \infty$ for $f\to f_c^-$.
  Data correspond to simulations of a system size $L=3360$.}
 \end{figure}
\fi

\iffigures
\begin{figure*}
  \centering
 \includegraphics[width=0.45\textwidth]{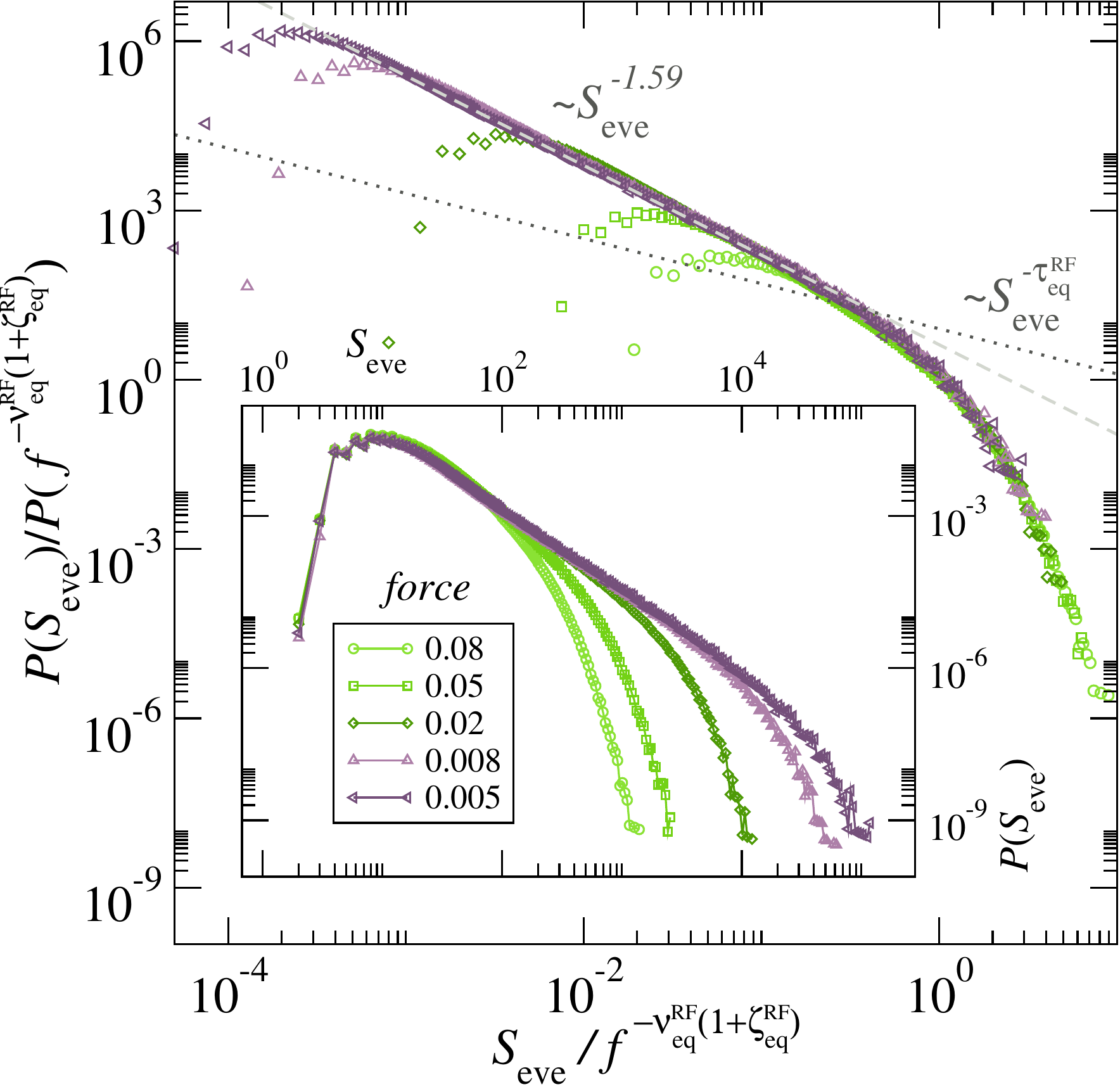}
  \hspace{0.2cm}
 \includegraphics[width=0.45\textwidth]{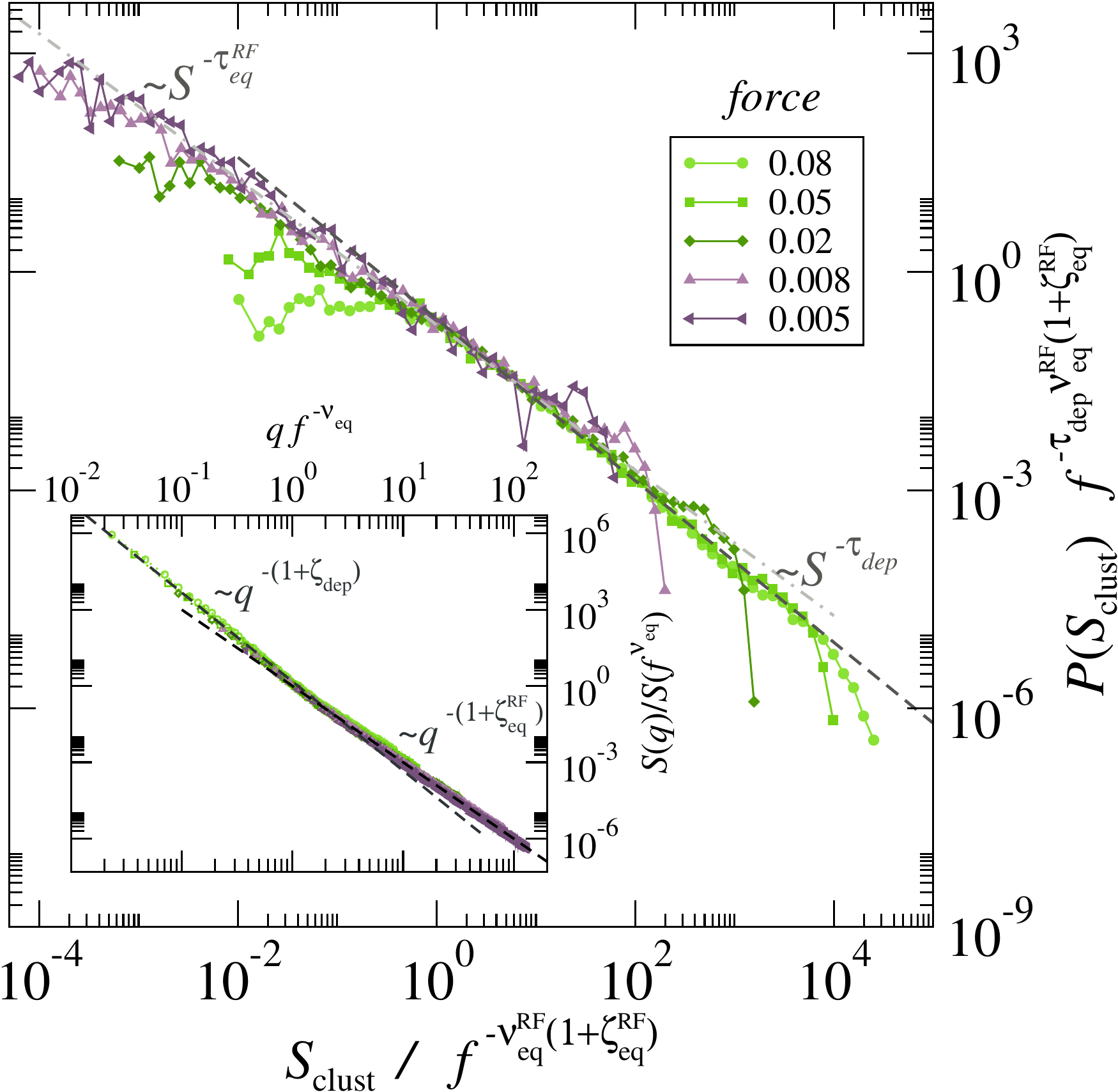} 
  \caption{
   \textit{Event and cluster size distributions for the Random Field disorder case~--} 
   On the left: Distribution of creep events in presence of RF disorder for different forces.  
   The inset shows the $P(\Seve)$ as a function of $\Seve$ while the main panel
   shows the collapse obtained using the rescaled variables $P(\Seve)/P(f^{-\nueqRF(1+\zetaeqRF)})$
   and $\Seve/f^{-\nueqRF(1+\zetaeqRF)}$. The collapse 
   confirms the creep scaling for the cutoff $S_c \sim \Lopt^{1+\zetaeqRF} \sim f^{-\nueqRF(1+\zetaeqRF)}$.
   On the right: Cluster size distribution for RF disorder and different forces.
   As in the RB case a crossover from equilibrium to depinning is observed when 
   we scale $\Sclus$ with $\Sopt \simeq f^{-\nueqRF(1+\zetaeqRF)}$.
   Notice that both the scaling size $\Sopt$ and the exponent of the equilibrium power-law regime
   ($\tau^{RF}_{{\text{eq}}}=1$) have changed respect to the ones of the RB case and still the scaling holds.
   Data correspond to a system of size $L=3360$.
  }\label{fig:PS_RF}
 \end{figure*}
\fi

The fact that RB and RF disorders have the same depinning 
but different equilibrium exponents has 
visible consequences in the creep event size distributions.
In fact, as can be seen
in Fig.~\ref{fig:PS_RF}-left, and comparing with 
Fig.~\ref{fig:PofSeve},
individual events decays faster for RF than for RB disorder. 
On the other hand in Fig.~\ref{fig:PS_RF}-right we show that the distribution 
of cluster sizes decays, for large sizes, as a power law with an exponent 
$\taudep \approx 2-2/(1+\zetadep) \approx 1.11$,
in accordance with Fig.~\ref{fig:PofSclus} and the fact that depinning
exponents are identical for RB and RF.
For sizes smaller that $\Sopt$ however, the RF cluster distribution
in Fig.~\ref{fig:PS_RF}-right is better described by a power law decay with exponent
$\taueqRF \approx 2-2/(1+\zetaeqRF) \approx 1$, 
which is visually distinguishable from the exponent $\taueq \approx 4/5$ of the RB case
in Fig.\ref{fig:PofSclus}.
As for the case of RB disorder, the structure factor $S(q)$ also accompanies
this length crossover, showing a geometrical change at wave-vector $q_c \simeq 1/\Lopt(f)$,
as displayed in the inset of Fig.~\ref{fig:PS_RF}-right.
But notice now that $\Lopt(f) \sim f^{-\nueqRF}$, with a different exponent.
Consistently, the geometrical crossover is now from the unique depinning
roughness $\zetadep$ at large scales, to the RF equilibrium roughness $\zetaeqRF$ at small ones.


The good collapse of the curves in Fig.~\ref{fig:PS_RF}-left under
the rescaling $\Seve/f^{-\nueqRF(1+\zetaeqRF)}$ shows that $ f^{-\nueqRF(1+\zetaeqRF)}$
controls the cutoff of the distribution, as it was shown for the RB case in the article.
Another way to systematically access the cutoff 
$S_c$ of a distribution of the form (\ref{eq:PSeve})
is to look at the higher moments $S_c \approx S_n \equiv \langle S^n\rangle/\langle S^{n-1}\rangle$
with $n>\tau$.

In Fig.~\ref{fig:cutoff} we show this ratio for $n=4$ both for RB and RF.
Further, for each type of disorder we show the value obtained considering
only the activated part $\Sact$ of the event (recall the algorithm description
above) and the total area of the event $\Seve$ (which is the activated plus the
deterministic part).
As can be seen comparing with the dashed lines representing $\Sopt \sim f^{-\nueq(1+\zetaeq)}$
(compatible with the celebrated $\Lopt \sim f^{-\nueq}$),
at small forces one recovers the creep scaling for the high order moment
(equivalently, for the cutoff of the distribution) $S_n \sim f^{-\nueq(1+\zetaeq)}$, be it RB or RF.
At large forces, $\Sact$ and the total area $\Seve$ are significantly different due to
the divergence of the deterministic relaxation after activation as we approach $\fc$.

In fact, $\Seve$ is defined as a combination of an activated move of size
$\Sact$, which diverges for $f \to 0$, plus a deterministic relaxation
diverging at $f \to \fc$.
Therefore, $\Seve$ diverges at both critical points while taking
a minimal value at intermediate forces.
This, translated to $\Leve$, accompanies the non-monotonic behavior of
the MSD discussed before.

In summary, the observations made in this section further confirm that 
creep motion is generically described by both equilibrium 
and depinning exponents regardless of the RB or RF nature of the disorder. 

\section{Quantitative estimations for the case of P\MakeLowercase{t/}C\MakeLowercase{o/}P\MakeLowercase{t} thin films}
\label{sec:experiments}

We believe that the clustering of creep events  
can be observed in experiments with the current apparatus and magneto-optical techniques 
which are able 
to directly visualize the interface motion. 
Indeed, using this technique, Repain \textit{et al.} \cite{RBJFM04} 
already observed, in a He-ion–irradiated Pt/Co(0.5 nm)/Pt
ultrathin film, small correlated events in the creep regime whose characteristic 
size increased with lowering the field, in good qualitative agreenment with 
our predictions. 
More Recently, Gorchon et al.~\cite{GBFJKG14} studied field-driven one dimensional
domain walls in ultrathin Pt(0.35nm)/Co(0.45nm)/Pt(0.35nm) magnetic films
with perpendicular anisotropy, using magneto-optical Kerr microscopy.
By fitting the velocity force characteristics in the creep and depinning regimes, 
they determined a critical depinning field $H_{\dep} \approx 1000\;$Oe and a
characteristic energy scale $k_B T_{dep} \approx 2000\;K$ at room temperature ($T=300\;K$). 
With these values it is possible to evaluate the  linear size of the event cut-off, $\Lopt$ 
and the associated displacement, $\hopt$, using the standard 
assumptions of weak pinning~\cite{larkin1979pinning,nattermann1990interface,demery2014microstructural}:
\begin{eqnarray}\label{Lopt_hopt}
  \Lopt =
  \Lc (H_{dep}/H)^\nueq 
  \approx 40  nm (H_{dep}/H)^{0.75}, \\
  \hopt =
  w_c (\Lopt/\Lc)^\zetaeq 
  \approx 20nm (H_{dep}/{H})^{0.5} .
\end{eqnarray}
Here  we used the known values for the 1d $\nueq$ and $\zetaeq$, from Table~\ref{tab:exponents}.
The microscopic length $L_c$, named Larkin  length, can be evaluated~\cite{larkin1979pinning,nattermann1990interface} as 
$L_c = (k_B T_{dep})/[(M_s H_{dep} \delta)w_c]\approx 0.04\; \mu m$, 
where $w_c\approx 20\;\mu m$ is of the order of the domain wall width, 
$\delta\approx 0.45\;nm$ is the thickness of the sample and
$M_s \approx 800 \;erg/G.cm^3$ the saturation magnetization.
Interestingly this implies that the two dimensional shape 
of creep events become increasingly compressed in the direction 
of motion as we reduce $H$, the aspect ratio scaling as 
$\Lopt/\hopt \approx 2 (H_{dep}/H)^{0.25}$
(very thin and elongated avalanches).
On the other hand, the area distribution cut-off of these events 
is predicted to scale as 
\begin{equation}
\Sopt = 
w_c L_c (\Lopt/\Lc)^{\zetaeq+1} 
\approx 800 \;nm^2 (H_{dep}/H)^{1.25}
\end{equation}
For a given field, detecting individual creep events of such sizes 
is mainly limited by the spatial resolution of the imaging technique.
Using a spatial resolution of $1\;\mu m$,  which is typical for 
magneto-optical setups, 
plugging in the measured $H_{dep} \approx 1000\;Oe$ and
asking for $\Lopt, \hopt > 1 \;\mu m$, 
we get the condition $H\lesssim 0.4 \;{Oe}$ at room temperature.
For such fields, 
$\Sopt = \hopt \times \Lopt \lesssim 40 \;\mu m \times 1\; \mu m = 40 \;\mu m^2$ 
and the average domain wall velocity drops below $1nm/s$,
as is found by extrapolating from the creep law.
At such small velocities, which are in principle still measurable
(e.g. $v_{min}=0.1\;nm/s$ in~\cite{MJMCF07}), the ``granularity''
of creep events should become observable, opening the possibility
to test our predictions of spatio-temporal correlations and
non-trivial distributions for them.

It is worth mentioning that $\Lopt$ was also estimated independently
in Ta(5.0-nm)/Pt(2.5-nm)/Co90Fe10 (0.3-nm)/Pt(1.0-nm) 
film wires with perpendicular magnetic anisotropy~\cite{KimNature2009},
with a completely different method.
Once more, $\Lopt$ was shown to scale as predicted for the RB 1$d$ model
with short-range elasticity $\Lopt\sim H^{-0.75}$.
This was achieved by observing the onset of finite effects in the velocity 
force characteristics at the creep regime, as the wire width $w$ was
physically reduced down to $w\sim \Lopt(H)$ and below. 
For these samples a field of $H=16\;Oe$ gives $\Lopt \approx 0.16\;\mu m$, 
remarkably in good agreement with the above estimate for the $Pt/Co/Pt$ film.
Interestingly enough, our results imply that at low fields, such that $\Lopt > w$,
the velocity is controlled by a power law distribution of thermally activated 
events with a size cut-off $w$.
This is consistent with the results reported in~\cite{KimNature2009}
showing the scaling of the velocity with $w$ in place of $\Lopt$.
Moreover, our findings also predict that the so-called ``zero dimensional''
regime in~\cite{KimNature2009} has actually contributions from 
power-law distributed correlated events with a size smaller than $w$.

\section{Movie}
Accompanying this manuscript we provide a movie illustrating the spatio-temporal 
patterns in ultra slow creep dynamics as obtained directly from our simulations
for a small $L=512$ system.
The movie displays the sequence of creep events in space and the corresponding
activity maps comparing with depinning avalanches. 


\end{document}